\documentclass[twocolumn]{aastex63}
\usepackage{booktabs}
\usepackage{amsmath}
\usepackage{threeparttable}

\shorttitle{Does the HCN/CO ratio trace the star-forming fraction of gas? I.}
\shortauthors{Bemis et al.}

\graphicspath{{Figs/}}

\begin{document}

\title{Does the HCN/CO ratio trace the star-forming fraction of gas?\\
I. A comparison with analytical models of star formation.}

\correspondingauthor{Ashley R. Bemis}
\email{bemis@strw.leidenuniv.nl}

\author[0000-0003-0618-8473]{Ashley R. Bemis}
\affiliation{Leiden Observatory,
Leiden University, 
PO Box 9513, 
2300 RA Leiden, 
The Netherlands}
\affiliation{McMaster University,
1280 Main St W, Hamilton ON L8S 4M1,
Canada}

\author[0000-0001-5817-0991]{Christine D. Wilson}
\affiliation{McMaster University,
1280 Main St W, Hamilton ON L8S 4M1,
Canada}

\begin{abstract}

We use archival ALMA observations of the HCN and CO $J=1-0$  transitions, in addition to the radio continuum at 93 GHz, to assess the relationship between dense gas, star formation, and gas dynamics in ten, nearby (U)LIRGs and late-type galaxy centers. We frame our results in the context of turbulent and gravoturbulent models of star formation to assess if the HCN/CO ratio tracks the  gravitationally-bound, star-forming gas in molecular clouds ($f_\mathrm{grav}$) at sub-kpc scales in nearby galaxies. We confirm that the HCN/CO ratio is a tracer of gas above $n_\mathrm{SF}\approx10^{4.5}$ cm$^{-3}$, but the sub-kpc variations in HCN/CO do not universally track $f_\mathrm{grav}$. {We find strong evidence for the use of varying star formation density threshold models, which are able to reproduce trends observed in $t_\mathrm{dep}$ and $\epsilon_\mathrm{ff}$ that fixed threshold models cannot. Composite lognormal and powerlaw models outperform pure lognormal models in reproducing the observed trends, even when using a fixed powerlaw slope. The ability of the composite models to better reproduce star formation properties of the gas provides additional indirect evidence that the star formation efficiency per free-fall time is proportional to the fraction of gravitationally-bound gas.}

\end{abstract}


\section{Introduction} \label{sec:intro}

A current challenge for understanding star formation in molecular clouds is determining the fraction of gas that is converted into stars over a cloud's lifetime. Observations show that sites of star formation are primarily in regions of dense molecular gas in Milky Way clouds \citep{Lada:1991a,Lada:1991b,Helfer:1997a,Helfer:1997b}, and these regions are confined to $\sim0.1$ pc scales within larger molecular structures in the form of clumps or filaments \citep{Andre:2016}. Within these structures of gas, it is the fraction of gravitationally-bound gas that goes on to form stars. Analytical models of star formation rely on estimates of the self-gravitating gas fraction, $f_\mathrm{grav}$, to then predict the star formation rate (SFR, e.g. \citealt{Krumholz:2005,Hennebelle:2011,Padoan:2011,Federrath:2012,Burkhart:2019}), which makes $f_\mathrm{grav}$ an important parameter to constrain observationally.
\par
Extragalactic observations rely on molecular transitions with high critical densities, $n_\mathrm{crit}\gtrsim10^4$ cm$^{-3}$ to gain information on the dense gas in other galaxies. The most commonly-used dense molecular gas tracer in extragalactic studies is the HCN $J=1-0$ transition \citep{Gao:2004a,Gao:2004b}.  Under the common assumption that the total emissivity of HCN  traces the dense gas mass, $I_\mathrm{HCN}\propto \Sigma_\mathrm{dense}$, then the ratio of the HCN and CO emissivities, $I_\mathrm{HCN}/I_\mathrm{CO}$, is proportional to the fraction of molecular gas in the dense phase, $f_\mathrm{dense}$. If the dense gas mass traced by HCN is also self-gravitating, then this line ratio is a simple, observational method for estimating $f_\mathrm{grav}$. However, the Interstellar Medium (ISM) of galaxies resides at a range of densities $\lesssim1-\gg 10^8$ cm$^{-3}$, and molecular transitions are sensitive to a continuum of these densities, including some fraction below their critical density \citep{Shirley:2015, Leroy:2017a}. Recent studies within the Milky Way have also shown that HCN may predominantly trace moderate gas densities \citep{Kauffmann:2017hcn_moderate}, rather than denser gas associated with star formation. The fraction of "CO-dark" or "CO-faint" gas may also contribute to variations in $I_\mathrm{HCN}/I_\mathrm{CO}$ \citep{ Grenier:2005, Wolfire:2010, Bolatto:2013}. On average in the Milky Way, $\sim30\%$ of molecular Hydrogen appears to be "CO-faint" \citep{Pineda:2013, Langer:2014}. This fraction is higher in lower-metallicity gas \citep{Pineda:2013,Jameson:2018,Chevance:2020} and appears to be more significant for clouds at lower masses \citep{Grenier:2005,Bolatto:2013} and at higher radii in the Milky Way \citep{Pineda:2013, Langer:2014, Chevance:2020}. This would likely result in overestimates of $f_\mathrm{dense}$ in lower mass clouds, assuming HCN emission is not affected in the same way. This is likely less significant in more extreme systems (such as U/LIRGs) or galaxy centers, such as those studied in this paper. 
\par
The dynamics of the gas in the ISM are set by a combination of gravity, turbulence, and magnetic fields, and these processes act together to set the spatial structure of gas in ISM clouds. These processes also likely play a role in setting the star formation properties of the ISM. One type of analytical model, gravoturbulent models of star formation, aims to predict both the observed structure of  gas clouds and their star formation properties. In particular, gravoturbulent models of star formation predict the shape of the gas volume density Probability Distribution Function ($n-$PDF) and the star formation efficiency ($\epsilon_\mathrm{ff}$) over a free-fall time ($t_\mathrm{ff}$)  \footnote{\citet{Brunt:2010} find that the two-dimensional column density distribution is a compressed version of the three-dimensional (volume) density distribution. \citet{Burkhart:2012} provide an analytical framework for connecting these two distributions in the case that both contain lognormal components, which we adopt in this paper.}\citep{Krumholz:2005,Padoan:2011,Hennebelle:2011,Federrath:2012,Burkhart:2018,Imara:2016}. \footnote{Free-fall time is treated differently depending on the framework used, and in reality it must be a reflection of multiple free-fall times from the array of gas densities that span the $n-$PDF.}. Studies have found that the \textit{column} density PDF of the diffuse ($n<1$ cm$^{-3}$) component of gas in the Milky Way and M33 is consistent with a lognormal PDF \citep[cf.][]{Berkhuijsen:2008,Hill:2008,Tabatabaei:2008, Burkhart:2015c}. The seminal analytical work by \citet{Vazquez-Semadeni:1994} showed that if the turbulent ISM  develops a series of isothermal and interacting supersonic shocks, the gas would naturally follow a lognormal PDF (cf. \citealt{Vazquez-Semadeni:1994, Padoan:1997,Scalo:1998,Nordlund:1999}. In this picture, the shocks amplify each other via a turbulent cascade of energy, and this multiplicative process results in the gas density PDF taking on a lognormal shape (cf. \citealt{Vazquez-Semadeni:1994, Padoan:1997,Scalo:1998,Nordlund:1999,Chen:2018}).
\par
Observations of molecular regions of the ISM reveal that the gas \textit{column} density PDF takes on a different form at high densities. The highest-density regions within more-evolved molecular clouds contribute a powerlaw tail to the gas column density PDF (cf. Chen et al. 2018), with some cloud PDFs being almost entirely powerlaw \citep[e.g.][]{Kainulainen:2009,Schneider:2013,Lombardi:2015,Schneider:2015a,Schneider:2016,Alves:2017}. This powerlaw has also been observed in simulations that develop self-gravitating gas (cf. \citealt{Ballesteros-Paredes:2011a,Collins:2012,Schneider:2015a,Burkhart:2017,Padoan:2017}. These results strongly suggest that the gas density PDF in a star-forming molecular cloud is likely a combination of a lognormal and powerlaw shape, and that the powerlaw tail is potentially the result of gas becoming self-gravitating. The fraction of gas within this powerlaw tail would then be the self-gravitating gas fraction, $f_\mathrm{grav}$. The density at which the $n-$PDF transitions from a lognormal to powerlaw shape would then represent when gas becomes available to star formation, $n_\mathrm{SF}$. The fraction of dense gas mass above $n_\mathrm{SF}$ ($f_\mathrm{grav}$), relative to the total mass of a star-forming cloud is then:
\begin{equation} \label{eq:fgrav}
    f_\mathrm{grav}=\frac{M(n>n_\mathrm{SF})}{M}
\end{equation}
\par
Turbulent models of star formation estimate $\epsilon_\mathrm{ff}$ by integrating over a purely lognormal gas density PDF, also above a gas threshold density (cf. \citealt{Krumholz:2005,Hennebelle:2011,Padoan:2011,Federrath:2012}). This threshold density is also meant to capture when gas becomes self-gravitating in the ISM, so that the fraction of gas above this threshold is $f_\mathrm{grav}$ in these models. However, lognormal-only models fail to explain observed variations in $\epsilon_\mathrm{ff}$ and mach number, $\mathcal{M}$, seen in some galaxies \citep[e.g.][]{Leroy:2017b}. For example, Giant Molecular Clouds (GMCs) in M51 show a weak anti-correlation between velocity dispersion (which is proportional to $\mathcal{M}$, see Eq. \ref{eq:mach}) and the star formation efficiency of gas per free-fall time, $\epsilon_\mathrm{ff}$.\footnote{We briefly discuss the observational uncertainties associated with estimates of $\mathcal{M}$, which depend on measurements of velocity dispersion, in \S \ref{sec:scale_dep}.} \citet{Leroy:2017a} argue that this anti-correlation may reflect differences in the dynamical state of their clouds with galactocentric radius.
\par
\citet{Burkhart:2019} show that including a powerlaw tail in the gas volume density PDF reproduces the observed variations in $\epsilon_\mathrm{ff}$ in M51, without requiring changes in the dynamical state of the clouds and without explicitly setting $n_\mathrm{SF}$. They find a slight anti-correlation between $\epsilon_\mathrm{ff}$ and $\mathcal{M}$ for virialized clouds ($\alpha_\mathrm{vir}\approx1$). This anti-correlation coincides with an increasing depletion time with $\mathcal{M}$, in agreement with the findings from PAWS \citep{Leroy:2017b}. $\epsilon_{\mathrm{ff}}\propto f_\mathrm{grav}$ implies $f_\mathrm{grav}$ may also anticorrelate with $\mathcal{M}$. Thus, without needing to explicitly set $n_\mathrm{SF}$, it is ultimately a decrease in $f_\mathrm{grav}$ with respect to increasing $\mathcal{M}$ that leads to the anti-correlation between $\sigma_\mathrm{rv}$ and $\epsilon_\mathrm{ff}$ \citep{Burkhart:2019}.  However, in starbursts where $\mathcal{M}$ is higher, we see higher $\epsilon_\mathrm{ff}$, and shorter $t_\mathrm{dep}$, on average \citep{Wilson:2019}. Starbursts typically have enhanced HCN/CO ratios in addition to shorter depletion time, $t_\mathrm{dep}$ \citep{Kennicutt:2021}.  If $I_\mathrm{HCN}/I_\mathrm{CO}\propto f_\mathrm{dense}$ and $f_\mathrm{dense}\propto f_\mathrm{grav}$, then the result from BM19 appears contrary to what is observed in starbursts.
\par
A potential explanation for these differences may be differences in the timescale for star formation, which may be set by the environment that a gas cloud is immersed in. The star formation law can be written as (\citealt{Krumholz:2012}):
\begin{equation} \label{eq:KDMlaw}
    t_\mathrm{ff} \Sigma_\mathrm{SFR} = \epsilon_\mathrm{ff} \Sigma_\mathrm{gas}
\end{equation}
where $\Sigma_\mathrm{gas}$ is the gas surface density, $\Sigma_\mathrm{SFR}$ is the star formation rate surface density, and $t_\mathrm{ff}$ is the free-fall time and is set by the self-gravity of a cloud. 
\par
\citet{Burkhart:2019} demonstrate the connection between $f_\mathrm{grav}$ and the instantaneous efficiency of the gas, $\epsilon_\mathrm{inst}\approx\epsilon_0\,f_\mathrm{grav}$, that reflects both the local efficiency, $\epsilon_0$ (set by e.g. stellar feedback), and $f_\mathrm{grav}$.  This local efficiency may correlate with its observational analog, the star formation efficiency per free-fall time, $\epsilon_\mathrm{ff}$ \citep{Krumholz:2005,Lee:2016}. Furthermore, if turbulence plays a significant role in setting $n_\mathrm{SF}$, then we may find a correlation between $\epsilon_\mathrm{ff}$ and observed velocity dispersions of gas, and turbulent pressure. Turbulent models of the ISM predict a dependence of $\sigma_\mathrm{n/n_0}$, the density variance of the volume density PDF ($n-$PDF) on the sonic mach number, $\mathcal{M}=\sigma_{v,3D}/c_s$\footnote{ The sound speed is given by $c_s = \sqrt{k T_\mathrm{kin}/\mu m_\mathrm{H}}$, $T_\mathrm{kin}$ is the gas kinetic temperature, $\mu=2.33$ \citep{Kauffmann:2008} is the mean molecular weight, and $m_\mathrm{H}$ is the Hydrogen mass. $\sigma_{v,3D}$ is the three-dimensional velocity dispersion and is related to the one-dimensional velocity dispersion via $\sigma_\mathrm{3D,v}=\sqrt{3}\sigma_\mathrm{v}$.}, within individual star-forming clouds. \citet{Lada:1994} found a correlation between the extinction, $A_\mathrm{V}$, in the Dark Cloud IC 5126 and the standard deviation of $A_\mathrm{V}$, with extinction increasing with dispersion \citep{Goodman:2009}. \citet{KainulainenTan:2013} find a correlation between measurements of the velocity dispersion from $^{12}$CO and $^{13}$CO and the density contrast ($N/N_0$) of column density PDFs ($N-$PDFs) derived from IR data in several Milky Way clouds. Combined, these correlations imply that the CO velocity dispersion may be sensitive to the density variance of the $N-$PDF, $\sigma_\mathrm{N/N_0}$ (where $\sigma_{N/N_0} \propto \mathrm{ln}(N/N_0)$), and therefore a probe of the ISM physics. 
\par
In external galaxies where resolution is limited, molecular line ratios are an additional tool for assessing $n-$PDF shape. \citet{Shirley:2015} showed that molecular transitions have an emissivity\footnote{Emissivity, which describes the emission per mass surface density, is effectively the inverse of a molecular line conversion factor.} that extends over a range of gas densities, including a significant amount of emission at densities below the critical density associated with that transition\footnote{Here the critical density is the density at which collisional interactions balance instantaneous de-excitation of a particular molecular transition \citep{DraineBook:2010}}. To determine if molecular line ratios stay sensitive to $n-$PDF shape, \citet{Leroy:2017a} model molecular line emissivities and explore a range of $n-$PDF shapes. They find that dense gas tracers (such as HCN and HCO$^+$) are more sensitive to changes in the shape of the $n-$PDF than lower-density tracers like CO.  Combining molecular line ratios of dense gas tracers with information on kinematics is therefore a promising tool for assessing $n-$PDF information of clouds in external galaxies.
\par
To assess the relationship between the HCN/CO ratio and $f_\mathrm{grav}$, we look to more extreme star-forming environments in which  turbulence is stronger (e.g. mergers, starbursts, (U)LIRGs, barred galaxies). We study a sample of 10 (U)LIRGs and disk galaxy centers that have archival CO, HCN, CN, and HCO$^+$ $J=1-0$ data, in addition to 93 GHz radio continuum. In this paper, we focus on general trends of the HCN/CO ratio, the star formation rate surface density, and $\epsilon_\mathrm{ff}$, and we compare these modelled trends with the observed trends in our sample. We use the EMPIRE  sample of galaxies \citep{Jimenez_Donaire:2019} as a comparison, which predominantly targets normal regions of star formation within galaxy disks.

\section{Data and Sample} \label{sec:data}

Our sample consists of ten nearby ($z<0.03$) galaxies, including the dense centers of five disk galaxies and five mergers and (U)LIRGs. We list these galaxies and their basic properties in Table  \ref{tab:source_data_props}. Four of the five disk galaxies in our sample are also barred. For each galaxy, we image archival ALMA data of the HCN, CN, CO, and HCO$^+$ $J=1-0$ transitions, in addition to the radio continuum emission at 93 GHz. The data are $uv-$matched and tapered to a common beam for each individual galaxy. The sample selection and data reduction process are presented in \citet{Wilson:2022} in detail, except for NGC 4038/9, NGC 1808, and NGC 3351, which have been reduced and imaged separately at higher velocity resolutions to be included in this analysis.

\begin{table*}[tb]
	\centering
	\caption{Spatial and spectral resolutions of the data for each galaxy.} \label{tab:source_data_props}
	\resizebox{\textwidth}{!}{%
	\begin{threeparttable}
	\begin{tabular}{lcccccccccc}
	\toprule
	Galaxy & Beam   &   Distance$^a$ &   Scale  &   $\delta v$ &   $i^b$    &   $z$	&	AGN	&	Bar &   Interacting &   Classification   \\
          &	$('')$  &   (Mpc)   &   (pc/beam)   &      (km s$^{-1}$) &   ($^\circ$) \\ \midrule
M83                 &   2.10    &   4.7   &   48  &   10    &   24      &  0.00171   &   --  &   Y   &--&   SB \\
Circinus            &   3.00    &   4.2   &   61  &   20    &   66      &  0.00145  &   Sy 2    &--&--&   SB   \\
NGC 3351            &   3.45    &   9.3   &  156  &   10   &   45.1    &  0.00260 &   --   &   Y   &N&   SB    \\
NGC 3627            &   4.15    &   9.4   &  189  &   20    &   56.5    &  0.00243  &  LINER/Sy 2  &   Y   &Y&   Post-SB  \\
NGC 1808            &   3.75    &   7.8   &  142  &   10    &   57      &  0.00322  &   Sy 2    &   Y   &--&   SB \\
NGC 7469            &   0.95    &   66.4    &   306    &   20   &   45    &  0.01632  &   Sy 1    &--&   Y   & LIRG \\
NGC 3256            &   2.20    &   44  &  469    &   27.5 &   --      &  0.00935  &   South Nucleus    &N&   Y   & LIRG  \\
NGC 4038            &   5.00    &   22  &  110   &   5.2  &   --       &  0.00569  &   --   &N&   Y   &   SB \\
IRAS 13120-5453     &   1.10    &   134 & 715 &   20  &   --          &  0.02076  &   Sy 2   &N& --  &   ULIRG  \\
VV114               &   2.30    &   81  &  903    &   20   &   --      &  0.02007  &   East Nucleus    &N&   Y   &   LIRG \\ \bottomrule
	\end{tabular}
	\begin{tablenotes}
	\item[a]  Distances from \citet{Wilson:2022}
	\item[b]  M83: \citet{Tilanus:1993}; Circinus: \citet{Jarrett:2003}; NGC 3351 \& NGC 3627: \citet{Sun:2020}; NGC 1808: \citet{Salak:2019} 
	\item[*] Distances and redshifts are also listed, which are used to determine the physical scale (in pc) per pixel, and to convert measured flux to luminosities. Redshifts are taken from the NASA/IPAC Extragalactic Database (NED). Inclination angles are taken from the papers listed below the table. Distances are the same as in \citet{Wilson:2022}, except for NGC 3627. We use the distance for NGC 3627 taken from \citet{Jimenez_Donaire:2019} for consistency when comparing with the EMPIRE data. We also list the presence/absence of an AGN, bar, or interaction with another galaxy. These classifications are taken from NED.
	\end{tablenotes}
	\end{threeparttable}}
\end{table*}

\subsection{Moment Maps}
We produce maps of integrated intensity and velocity dispersion of the CO and HCN $J=1-0$ transitions using the Astropy Spectral Cube package \citep{Ginsburg:2019}. We implement a masking method similar to that in \citet{Sun:2018} and summarize the masking method here:
\begin{enumerate}
    \item The r.m.s. noise is estimated in each channel of each datacube using Median Absolute Standard Deviation.
    \item Peaks of emission with a signal-to-noise ratio (SNR) of at least five across two channels are identified within each datacube.
    \item Masks are expanded around these peaks down to channels with emission at a SNR of three.
    \item Emission from regions smaller than a beam area are masked.
\end{enumerate}
We calculate the uncertainty in integrated intensity and velocity dispersion using Eqs. \ref{eq:unc_mom0} and \ref{eq:unc_mom2}, respectively. We require that all pixels have $\mathrm{SNR}>3$ in integrated intensity in addition to a $\mathrm{SNR}>2$ in velocity dispersion.

\subsection{Molecular Gas Surface Densities}

One of the main goals of this study is to assess variations in the fraction of gas in the dense phase as traced by the HCN and CO molecular line luminosities. Our analysis allows us to estimate trends in the HCN and CO luminosity-to-mass conversion factors (cf. \citealt{Bolatto:2013}), which we define as:
\begin{align} 
    \alpha_\mathrm{mol} &= \frac{\Sigma_\mathrm{mol}}{I_\mathrm{mol}}\qquad[\mathrm{M}_\odot\,\mathrm{pc}^2\, (\text{K km s}^{-1})^{-1}]
\end{align}
where $\Sigma_\mathrm{mol}$ is the mass surface density of the molecular gas, including Helium,  $I_\mathrm{mol}=L_\mathrm{mol}/A_\mathrm{pix}$ is the intensity in units of K km s$^{-1}$, and $L_\mathrm{mol}$ is total luminosity over the physical area of a pixel, $A_\mathrm{pix}$. To calculate the molecular gas mass surface density, $\Sigma_\mathrm{mol}$, we use:
\begin{equation} \label{eq:Sigma_mol}
    \Sigma_\mathrm{mol} = \alpha_\mathrm{mol}\ I_\mathrm{mol}\ \mathrm{cos}(i).
\end{equation}
\noindent We apply inclination angles only to the disk galaxies in this sample, as inclination angles are typically uncertain in mergers and (U)LIRGs. Thus, the uncorrected measurements of galaxies with non-zero inclinations will result in overestimates of $\Sigma_\mathrm{mol}$ and other surface densities.
\par
For ease of comparison with other studies our fiducial value is $\alpha_\mathrm{CO}=1.1$ [M$_\odot$ (K km s$^{-1}$ pc$^2$)$^{-1}$] for our sample of galaxies, the (U)LIRG value including Helium \citep{Downes:1993}, which is $\sim$4 times lower than the Milky Way value. This lower value is motivated by evidence that gas in these systems is subject to more extreme excitation mechanisms, e.g. higher temperatures and densities (cf. \citealt{Bolatto:2013,Downes:1993} and references therein). Additionally, the gas traced by CO in these systems often shows broad line widths, potentially reducing the opacity of the CO transition \citep{Bolatto:2013,Downes:1993}. \citet{Downes:1993} also suggest that CO may be subthermally excited ($T_\mathrm{ex} < T_\mathrm{kin}$) in starbursts and (U)LIRGs. We choose a fixed value of $\alpha_\mathrm{CO}=4.35$ [M$_\odot$ (K km s$^{-1}$ pc$^2$)$^{-1}$] for the EMPIRE sample of galaxies, since these are mostly disk galaxies and are likely more similar to the Milky Way than to starbursts. Since we already have NGC 3627 in our sample, we drop NGC 3627 from the EMPIRE data. 
\par
The HCN conversion factor is less certain. Historically, $\alpha_\mathrm{HCN} \approx 13.6$ [$\mathrm{M}_\odot$ (K km s$^{-1}$ pc$^2)^{-1}$] has been used \citep{Gao:2004a,Gao:2004b}, which is appropriate for a virialized cloud core with a mean density $n_\mathrm{H_2}\sim3\times10^4$ cm$^{-3}$ and brightness temperature $T_\mathrm{B}\sim35$ K (e.g., \citealt{Radford:1991a}, including Helium). This HCN conversion factor assumes that this molecular transition is optically thick, and that the gas it traces is in local thermodynamic equilibrium (LTE). In \citet{Bemis:2020} and Bemis \& Wilson, in prep. we explore variations in the relative values of $\alpha_\mathrm{HCN}$ and $\alpha_\mathrm{CO}$ using a non-LTE radiative transfer analysis. If all of the above assumptions are true, then the fraction of dense gas traced by HCN/CO is given by:
\begin{equation}
    f_\mathrm{dense} = \frac{\alpha_\mathrm{HCN}}{\alpha_\mathrm{CO}} \frac{I_\mathrm{HCN}}{I_\mathrm{CO}}.
\end{equation}
with $\alpha_\mathrm{HCN}/\alpha_\mathrm{CO}=3.2$ implied from the above discussion. We do this assuming that the physical effects impacting $\alpha_\mathrm{CO}$ will impact $\alpha_\mathrm{HCN}$ in a similar way. Simulations show that a slight reduction in $\alpha_\mathrm{HCN}$ does happens in presence of higher gas temperatures ($T_\mathrm{kin}=20$ K, cf. \citealt{Onus:2018}), similar to our expectation for $\alpha_\mathrm{CO}$. Furthermore, by using $\alpha_\mathrm{HCN}/\alpha_\mathrm{CO}=3.2$ our plots of $f_\mathrm{dense}$ are a direct reflection of the HCN/CO ratio. We distinguish this calculation of $f_\mathrm{dense}$ from analytical estimates of the fraction of dense, star-forming gas which is generally defined as the mass fraction above the density at which gas becomes self-gravitating, $f_\mathrm{grav}$.

\subsection{The Radio Continuum SFR}

We detect radio continuum emission at 93 GHz in all of our sources and use this as our SFR tracer. The radio continuum is a combination of thermal (T) free-free emission and non-thermal (NT) synchrotron emission from regions with massive star formation, spanning $\sim1-100$ GHz. At 93 GHz, we are in the regime where thermal free-free emission from young star-forming regions will likely dominate the radio continuum emission. Non-thermal emission is expected to contribute $\sim25\%$ to the radio continuum luminosity at this frequency (see \citealt{Wilson:2019,Murphy:2011}), assuming an electron temperature $T_\mathrm{e}\sim10^4$ K and non-thermal spectral index $\alpha_\mathrm{NT}\sim0.84$. This fraction will change if there are variations in either $T_\mathrm{e}$ or $\alpha_\mathrm{NT}$. 
We adopt fixed values for $\alpha_\mathrm{NT}=0.84$ and $T_\mathrm{e}=10^4$ K, and we use the composite calibration from \citet{Murphy:2011} which accounts for both thermal and non-thermal contributions to the SFR:

\vspace{1em}
\begin{multline} \label{eq:sfr_radio_both}
\left( \frac{\mathrm{SFR}_\nu}{M_\odot\ \mathrm{yr}^{-1}} \right) = \\
10^{-27} \Big[ 2.18\left( \frac{T_\mathrm{e}}{10^4\ \mathrm{K}} \right)^{0.45}\left( \frac{\nu}{\mathrm{GHz}} \right)^{-0.1} + \\
 15.1\left( \frac{\nu}{\mathrm{GHz}} \right)^{-\alpha^\mathrm{NT}} \Big]^{-1} \left( \frac{L_\nu}{\mathrm{erg\ s^{-1}\ Hz^{-1}}} \right)
\end{multline}
\vspace{1em}

At 93 GHz, the radio continuum emission from star-forming regions may overlap with the lower-frequency tail of the dust SED. \citet{Wilson:2019} estimate a $\sim10\%$ contribution from dust at 93 GHz for IRAS 13120-5453, NGC 3256, and NGC 7469, the three most IR-luminous galaxies in our sample. We adopt this correction factor of $\sim10\%$ for all sources, but we acknowledge that the emissivity of dust at 93 GHz may vary between our sources. We further mask pixels that may be contaminated with AGN emission (cf. Table \ref{tab:source_data_props}).
\par 
Variations in $T_\mathrm{e}$ and $\alpha_\mathrm{NT}$ will impact our SFR estimates from the radio continuum at 93 GHz. Electron temperatures typical of H\textsc{II} regions are $T_\mathrm{e}\sim5\times10^3-10^4$ K, which will produce a change of $\sim30\%$ \citep{Murphy:2011}. In contrast to this, $\alpha_\mathrm{NT}$ has been observed as low as $\sim0.5$, which would also give a change in luminosity $\sim30\%$ at 93 GHz. \citet{Wilson:2019} also find evidence of a significant fraction (up to $50\%$) of non-thermal emission in NGC 7469 at 93 GHz by comparing with an archival radio continuum map at 8 GHz. We adopt an uncertainty of $30\%$ for SFR estimates derived from the radio continuum.
\par
We compare the results of our sample with those of the EMPIRE survey \citep{Jimenez_Donaire:2019}, for which there are publicly-available single-dish (IRAM 30m) observations of HCN and CO.  We estimate $\Sigma_\mathrm{SFR}$ in EMPIRE galaxies using 24 $\mu$m IR maps from the \textit{Spitzer} Space Telescope. These IR data are convolved to a $15''$ Gaussian beam utilizing \citet{Aniano:2011} Gaussian kernels and further smoothed to a $33''$ Gaussian beam using \textsc{CASA} \citep{McMullin:2007}. Backgrounds are subtracted and SFRs are derived using the \citet{Rieke:2009} calibration.

\subsubsection{Star Formation Timescales and Efficiency}

 We use the star formation rate (SFR) and molecular gas surface densities to estimate the depletion time of the total (mol$ \,=\, $CO) and dense (mol$ \,=\, $HCN) molecular gas content:
\begin{align}
    t_\mathrm{dep} &= \frac{\Sigma_\mathrm{mol}}{\Sigma_\mathrm{SFR}}
\end{align}
where $\Sigma_\mathrm{mol}$ is estimated using Eq. \ref{eq:Sigma_mol}. 
To estimate the dimensionless star formation efficiency, we compare this depletion timescale with the free-fall timescale:
\begin{align}\label{eq:tff}
    t_\mathrm{ff} &= \sqrt{\frac{3\pi}{32 G \rho}}
\end{align}
where $\rho$ is a characteristic density of the gas associated with star formation. 
We calculate gas density assuming a fixed line of sight (LOS) depth via  $\rho\approx\Sigma_\mathrm{mol}/R$, where $R$ is equivalent to half of the `cloud' or gas depth along the line of sight (LOS). We do not know $R$, so we assume a fixed value of 100 pc for all galaxies when estimating $\rho$. We also explored estimating free fall time using the \citet{Krumholz:2012} prescriptions for $t_\mathrm{ff}$ in the GMC and Toomre regimes. We find little qualitative difference in the results between using $t_\mathrm{ff}$ as described above and the \citet{Krumholz:2012} prescription.
\par
The efficiency of the star-formation process is then estimated by comparing the observed depletion timescales with estimates of the free-fall time:
\begin{equation}
    \epsilon_\mathrm{ff}=\frac{t_\mathrm{ff}}{t_\mathrm{dep}}
\end{equation}
which is just the star formation law (Eq. \ref{eq:KDMlaw}) re-written in terms of depletion time.

\subsection{Velocity Dispersion}

We measure the 1-dimensional velocity dispersion of the molecular gas in our sources, $\sigma_v$, using the CO $J=1-0$ transition. The velocity dispersions, $\sigma_\mathrm{v,meas}$, are measured directly from moment 2 maps. We correct for broadening of the line due to the finite spectral resolution of our data using \citet{Rosolowsky:2006}:
\begin{equation}
    \sigma_v = \sqrt{\sigma_{v\mathrm{meas}}^2 - \frac{\delta{v}^2}{2\pi}}
\end{equation}
where  $\delta v$ is the channel resolution at which we image the data.  This value is then converted to the 3-dimensional velocity dispersion via $\sigma_{v,3D} = \sqrt{3}\sigma_v$. 

\subsubsection{Velocity Dispersion as a Tracer of Mach Number}

As previous studies have done for Milky Way clouds \citep{Kainulainen:2017}, we use CO linewidths as an indicator of the mach number of the gas in our galaxies:
\begin{equation}\label{eq:mach}
    \mathcal{M} = \frac{\sqrt{3} \sigma_\mathrm{v,1D}}{c_s},
\end{equation}
where $c_\mathrm{s}$ is the thermal sound speed and $\sigma_\mathrm{v,1D}$ is the observed one-dimensional velocity dispersion. There are limits to this approach, and the ability of the CO line to trace cloud turbulence may be limited by its optical depth \citep{Burkhart:2013,Goodman:2009}, and observed line width can include disk rotation or other large-scale motions, but alternative measures of gas kinematics in extragalactic clouds are absent.
\par
To estimate $c_s$, we consider an ideal gas equation of state ($P_\mathrm{th} = n k_\mathrm{B} T / \mu m_\mathrm{H}$),  such that the sound speed is $c_s = \sqrt{k_\mathrm{B} T / \mu m_\mathrm{H}}$, where $k_\mathrm{B}$, $T$, $\mu$, and $m_\mathrm{H}$ are the Boltzmann constant, gas kinetic temperature, the mass of a hydrogen atom, and the mean particle weight. For a gas where hydrogen is primarily in molecular form $\mu = 2.33$ (assuming cosmic abundances, \citealt{Kauffmann:2008}). For $\mu = 2.33$ and a temperature rage of $T = 10-100$ K, $c_s \approx 0.2-0.6$ km s$^{-1}$. The choice of $c_s$ can have a significant impact on the estimate of $\mathcal{M}$. We choose an intermediate value $c_s=0.4$ km s$^{-1}$, corresponding to $T\sim45$ K as our fiducial value.

\subsection{Uncertainties From a Multi-scale Sample} \label{sec:scale_dep}

The spatial resolution of the data in our sample span $\sim50-900$ pc. We consider how this range may impact our measurements from the context of a turbulent ISM, and how we can best interpret measured $\sigma_v$ and line ratios in our sources. The beam filling fraction, $\phi_\mathrm{ff,beam}$ is less of an uncertainty in this study, since the sources with the lowest resolution are (U)LIRGs and likely have filling fractions approaching unity, while the spiral galaxy centers have resolutions approaching cloud scales. This reduces the issue of variations in beam filling fractions from galaxy-to-galaxy. A significant source of uncertainty instead comes from variations in the relative filling fractions of the HCN/CO transitions:
\begin{equation}
    \Phi_\mathrm{HCN/CO} \approx \frac{r_\mathrm{HCN}^2}{r_\mathrm{CO}^2},
\end{equation}
\noindent where $r_\mathrm{HCN}$ and $r_\mathrm{CO}$ are the average radial extents of HCN and CO, respectively. We explore variations in the filling fraction in \citet{Bemis:2020} and Bemis \& Wilson, in prep. from a radiative transfer perspective.
\par
In addition to the uncertainty in $\Phi_\mathrm{HCN/CO}$, the velocity dispersion in a turbulent medium is scale-dependent ($\ell$) such that \citep{Larson:1981,Heyer:2009}:
\begin{equation} \label{eq:sigmav_turb}
    \sigma_v(\ell) = \sigma_{v,L} \left( \frac{\ell}{L} \right)^{p}
\end{equation}
\noindent where $p\sim0.33-0.5$ depends on the type of turbulence, (e.g. $p\approx0.33$ for Kolmogorov, \citealt{Larson:1981}), $L$ can be defined as the cloud diameter (or turbulent injection scale) and $\sigma_\mathrm{v,L}$ is the cloud dispersion at that scale. A sample of galaxies at different physical scales will therefore be affected by this scale dependence, such that velocity dispersions will be smaller at smaller scales and vice versa. However, this relationship should saturate at the turbulent injection scale, which may mitigate some of this uncertainty. If larger-scale cloud-cloud motions do not affect the velocity dispersions of CO, then this saturation should be detected, although there is still uncertainty in $L$. This means that lower-resolution observations may still be useful for assessing $\sigma_\mathrm{v}$.

\section{Model Framework: Gravoturbulent Models of Star Formation}
\label{sec:fgrav_theory}

We consider several prescriptions of $f_\mathrm{grav}$ in the context of analytical models of star formation, and compare the predictions of these models to measurements of the $I_\mathrm{HCN}/I_\mathrm{CO}$ ratio and the star formation properties of our galaxies.  We use Eq. \ref{eq:KDMlaw}, which is relevant to  gravoturbulent models of star formation (KMD12).
\par
$\epsilon_\mathrm{ff}$ is calculated by integrating over the star-forming portion of the density-weighted gas density Probability Distribution Function (PDF, e.g \citealt{Krumholz:2005,Padoan:2011,Hennebelle:2011,Federrath:2012}):
\begin{equation} \label{eq:sfr_ff}
    \mathrm{\epsilon}_\mathrm{ff} = \epsilon_0 \int_{n_\mathrm{SF}}^{\infty} \frac{t_\mathrm{ff} (n_0)}{t_\mathrm{ff} (n)} \frac{n}{n_0} p(n)\mathrm{d}n
\end{equation}
\noindent Here $p(n)$ is the cloud's volumetric density PDF, $n-$PDF, and $n_0$ is its mean density. The $n-$PDF characterizes the distribution of densities of \textit{all} gas within the cloud, including the diffuse atomic and molecular components. We briefly review specific terms within this equation individually.
\par
$p(n)$: Isothermal gas in the presence of supersonic turbulence naturally becomes distributed such that its $n-$PDF is roughly lognormal \citep{Vazquez-Semadeni:1994,Nordlund:1999,Wada:2001}. Earlier formalisms of gravoturbulent models adopt a purely lognormal form of $p(n)$ (e.g. \citealt{Krumholz:2005,Padoan:2011,Federrath:2012}). In terms of the logarithmic volume density $s=\mathrm{ln} (n/n_0)$, this is given by:
\begin{equation}\label{eq:pdf_lognormal}
    \mathrm{p}_s =
    N \frac{1}{\sqrt{2\pi \sigma_s^2}}\mathrm{exp}\left( -\frac{(s-s_0)^2}{2\sigma_s^2}\right)
\end{equation}
where $N$ is a normalization constant\footnote{We use $N$ given by \citet{Burkhart:2018}.}, $\sigma^2_s$ is the logarithmic density variance which depends on the underlying physics of the gas and sets the width of $p(n)$, and $s_0=-0.5\, \sigma_s^2$. Newer formalisms (i.e. \citealt{Burkhart:2018}) of gravoturbulent models of star formation suggest that the $n-$PDF may then evolve to include a high-density powerlaw tail which contains gas that is becoming gravitationally unstable to collapse. Observations of clouds in the Milky Way support a composite form of $p(n)$, where a powerlaw tail is clearly present in addition to a more-diffuse component of gas. \citet{Burkhart:2018} presents this as a piecewise function:
\begin{equation}\label{eq:pdf_piecewise}
    \mathrm{p}_s =
\begin{cases}
    N \frac{1}{\sqrt{2\pi \sigma_s^2}}\mathrm{exp}\left( -\frac{(s-s_0)^2}{2\sigma_s^2}\right), & s<s_t \\
    N C e^{-\alpha_\mathrm{PL} s}, & s>s_t
\end{cases}
\end{equation}
where $s_t$ is the logarithmic form of the transition density between the two components of the $n-$PDF, and $\alpha_\mathrm{PL}$ is the slope of the powerlaw tail. The factors $N$ and $C$ are normalization constants given by \citet{Burkhart:2018}.
\par
$\sigma_s$: Numerical studies have shown that if the turbulence is super-Alfv\'{e}nic, and the magnetic field ($B$) follows a powerlaw relationship with gas density, $B\propto n^{1/2}$, then the logarithmic density variance is given by \citep{Molina:2012,Federrath:2012}: 
\begin{equation} \label{eq:sigma_s_chap3}
    \sigma_s^2 = \mathrm{ln}\left(1+b^2 \mathcal{M}^2 \frac{\beta}{\beta+1} \right)
\end{equation}
where $\mathcal{M}$ is the mach number, $\beta = P_\mathrm{th}/P_\mathrm{mag}$ is the plasma beta (which characterizes the ratio of thermal pressure to magnetic pressure), and $b$ is the turbulent forcing parameter (which characterizes the relative amount of solenoidal $b=1$ or compressive $b=0.33$ turbulence within the gas). Without a clear prescription for $\beta$ and $b$ in our sample, we simply take $\beta\rightarrow\infty$, which assumes that $B=0$ G. We take an intermediate value $b=0.4$, which assumes that turbulence is a mixture of compressive and solenoidal forcing. Non-isothermal gas will add intermittency resulting in deviations from a lognormal shape of $p(n)$ \citep{Federrath:2015}. For the purpose of this work, we assume an underlying lognormal is a reasonable, intermediate approximation to part of the $n-$PDF \citep{Federrath:2015}.
\par
$n_\mathrm{SF}$:  Gravity overcomes supportive processes and begins the process of collapse above the density threshold, $n_\mathrm{SF}$. The processes competing with gravity include some combination of magnetic support, internal turbulent motions, external ISM pressures, etc. Any gas with densities higher than $n_\mathrm{SF}$ is then potentially gravitationally unstable and star-forming, so $n_\mathrm{SF}$ serves as the lower limit of the integration that determines $\epsilon_\mathrm{ff}$.
\par
$t_\mathrm{ff}(n_0)/t_\mathrm{ff} (n)$: Within the integral in Eq. \ref{eq:sfr_ff} integral is a free-fall time factor, which  converts the integrand into a dimensionless mass \textit{rate} equivalent to the mass per free-fall time \citep{Federrath:2012}. This time factor is treated differently depending on the analytical model being considered, e.g. single free-fall (SFF) time \citep{Krumholz:2005,Padoan:2011} versus multi-free-fall (MFF) time models \citep{Hennebelle:2011,Federrath:2012}, and these differences are summarized in \citet{Federrath:2012} and \citet{Burkhart:2018}. MFF models keep this factor in the integral, which predicts different rates of collapse for different densities, while SFF models take this factor out of the integral. Due to this difference, MFF models predict higher $\epsilon_\mathrm{ff}$.
\par
$\epsilon_0$: The prefactor in Eq. \ref{eq:sfr_ff}, $\epsilon_0$, is the \textit{local} dimensionless efficiency of star formation. $\epsilon_0$ depends on additional processes, such as the level of stellar feedback dispelling some of the gas that is already above $n_\mathrm{SF}$ \citep{Burkhart:2019}. The mass that does get converted into stars is then $M_*=\epsilon_0 M(n>n_\mathrm{SF})$.

\subsection{Different Formalisms of Gravoturbulent Models of Star Formation}
\citet{Federrath:2012} and \citet{Burkhart:2018} both provide analytical equations for estimating $\epsilon_\mathrm{ff}$ for different gravoturbulent formalisms which we adopt in this analysis. In this work, we focus on three analytical models which each have unique prescriptions for the density threshold and the shape of the $n-$PDF: 1. the \citet{Padoan:2011} formalism, 2. the \citet{Burkhart:2018} and \citet{Burkhart:2019} formalism, and 3. a fixed density threshold model, which is not explicitly formulated in previous works but is often referenced in the literature when interpreting the results of studies using dense gas tracers (e.g. \citealt{Gao:2004a,Gao:2004b}).
\par
We mainly refer to analytical equations from \citet{Burkhart:2018} for brevity, but also refer the reader to \citet{Federrath:2012} for another comprehensive summary of analytical models, as well as the original papers that have provided the basis of this work, e.g. \citet{Krumholz:2005,Padoan:2011,Hennebelle:2011}. We summarize the following important quantities of each of the models we use in this work: (a) the underlying $n-$PDF shape of the model, $p(n)$; (b) the equation used to estimate $\epsilon_\mathrm{ff}$; (c) the star formation threshold density, $n_\mathrm{SF}$; (d) and the method we use to determine $f_\mathrm{grav}$:
\begin{enumerate}
    \item The \citet{Padoan:2011} formalism:
    \begin{enumerate}
        \item These models have a lognormal (LN) $n-$PDF (see Eq. \ref{eq:pdf_lognormal}).
        \item \citet{Padoan:2011} estimate $\epsilon_\mathrm{ff}$ using (cf. Eq. (13) in \citealt{Burkhart:2018}):
        \begin{multline} \label{eq:PN11}
            \epsilon_\mathrm{ff,PN11} = \frac{\epsilon_0}{2} \left\{1 + \mathrm{erf}\left(\frac{\sigma_2^2 - 2\, s_\mathrm{thresh}}{\sqrt{8} \sigma_s } \right)\right\} \\
            \times \mathrm{exp}\left(s_\mathrm{thresh} / 2 \right)
        \end{multline}
        where $s_\mathrm{thresh}=\mathrm{exp}\left(n_\mathrm{SF}/n_0\right)$. \footnote{This formula is almost identical to the \citet{Krumholz:2005} formalism for $\epsilon_\mathrm{ff}$ if we remove the exponential factor containing $s_\mathrm{thresh}$.}
        \item The \citet{Padoan:2011} equation for $n_\mathrm{SF}$ that determines $s_\mathrm{thresh}$ in Eq. \ref{eq:PN11} is given by (cf. Eq. (11) in \citet{Burkhart:2018}):
        \begin{equation} \label{eq:ncrit_km05}
            n_\mathrm{SF} \approx 0.54\, n_0\,  \alpha_\mathrm{vir}\, \mathcal{M}^2\
        \end{equation}
        \noindent when we have taken the prefactor in their original equation to be $\phi=0.35$ and have neglected magnetic fields. The remaining dependence is then only with mean density, virial parameter ($\alpha_\mathrm{vir}$), and $\mathcal{M}$. \footnote{This equation is nearly identical to the \citet{Krumholz:2005} formalism, which has a prefactor $\pi/15$ (assuming $\phi_t=1$).}  We can also rewrite this equation in terms of the turbulent pressure, which we are able to estimate directly from our data \citep[cf.][]{Walker:2018}:
        \begin{equation} \label{eq:nthresh_pturb}
            n_\mathrm{SF} \approx0.36\, \alpha_\mathrm{vir} \frac{P_\mathrm{turb}}{k_B\, T_\mathrm{kin}}
        \end{equation}
        where $P_\mathrm{turb} \approx (3/2) \Sigma_\mathrm{mol} \sigma_{v}^2/R$ or $P_\mathrm{turb} \approx (3/2) n_0\sigma_{v}^2$ for the data and models, respectively. The direct scalings $n_\mathrm{SF}\propto \mathcal{M}^2$ and $n_\mathrm{SF}\propto P_\mathrm{turb}$ assume that turbulence acts as a supportive process to the gas. 
        The virial parameter, $\alpha_\mathrm{vir}$, is given by:
        \begin{equation}
            \alpha_\mathrm{vir} \approx \frac{5\sigma_v^2 R}{G\Sigma}
        \end{equation}
        \noindent and is the ratio of internal kinetic to gravitational energy in a cloud, $E_\mathrm{kin}$ and $E_\mathrm{grav}$.
        \item For these models, we perform a numerical integration of the $n-$PDF above $n_\mathrm{SF}$ (Eq. \ref{eq:ncrit_km05}) to calculate $f_\mathrm{grav}$. \footnote{$n_\mathrm{SF}$ is referred to as $n_\mathrm{crit}$ in \citet{Padoan:2011} and \citet{Burkhart:2018}, not to be confused with the critical density for a molecular transition.}.
    \end{enumerate}
    \item The \citet{Burkhart:2018} and \citet{Burkhart:2019} formalism:
    \begin{enumerate}
        \item These models have a piecewise (lognormal plus powerlaw, LN+PL) $n-$PDF (cf. Eq. \ref{eq:pdf_piecewise}).
        \item \citep{Burkhart:2018} calculate $\epsilon_\mathrm{ff}$ using a composite LN+PL $n-$PDF (cf. Eq. (27) in \citealt{Burkhart:2018}):
        \begin{multline}\label{eq:eff_BM18}
            \epsilon_\mathrm{ff,BM18} = \\
            N\mathrm{exp}\left(s_\mathrm{thresh} / 2 \right)     \frac{\epsilon_0}{2}\left\{ \mathrm{erf}\left(\frac{\sigma_2^2 - 2\, s_\mathrm{thresh}}{\sqrt{8} \sigma_s } \right)\right. -\\ 
            \mathrm{erf}\left(\frac{\sigma_2^2 - 2\, s_\mathrm{t}}{\sqrt{8} \sigma_s } \right) + \left.C\frac{\mathrm{exp(s_t(1-\alpha_\mathrm{PL}))}}{\alpha_\mathrm{PL} -1}\right\}
        \end{multline}
        where $s_t$ is the density at which the $n-$PDF transitions from a lognormal to power law shape, and $N$ is a normalization factor. We take $s_\mathrm{thresh}\equiv s_t$. 
        \item We adopt $s_t$ as the threshold density for the \citet{Burkhart:2018} formalism, which is given in \citet{Burkhart:2019} as:
        \begin{equation}\label{eq:s_t}
            s_t = (\alpha_\mathrm{PL} - 1/2)\sigma_s^2.
        \end{equation}
        In this formalism, $s_t$ is tied to the slope of the PL component of the $n$-PDF, as well as the width of the lognormal component.
        \item For the LN+PL models, $f_\mathrm{grav}$ is given by Eq. (20) in \citet{Burkhart:2019} (which is referred to as $f_\mathrm{dense}$ in their work). 
    \end{enumerate}
    \item Fixed density threshold models:
    \begin{enumerate}
        \item These models have a lognormal (LN) $n-$PDF (see Eq. \ref{eq:pdf_lognormal}).
        \item We again use Eq. (13) from \citet{Burkhart:2018} to calculate $\epsilon_\mathrm{ff}$.
        \item We use a fixed density threshold of $n_\mathrm{SF}=10^{4.5}$ cm$^{-3}$.
        \item We perform a numerical integration of the $n-$PDF above $n_\mathrm{SF}$ to calculate $f_\mathrm{grav}$.
    \end{enumerate}
\end{enumerate}

\noindent We hereafter refer to these three models as (1) LN PN11, (2) LN+PL B18, and (3) LN Fixed.

\subsection{Predictions of Analytical Models}
\label{sec:model_predictions}

Each of the models above has a unique prescription for $f_\mathrm{grav}$ depending on $n_\mathrm{SF}$ and the shape of the $n-$PDF, and this has an impact on their predictions of star formation. The simplest prescription for $f_\mathrm{grav}$ is that of fixed density-threshold models, which predict that $n_\mathrm{SF}\approx10^{4.5}$ cm$^{-3}$. In this context, the fraction of star-forming gas is any gas above this density. For purely lognormal $n-$PDFs, increases in the width of the $n-$PDF, $\sigma_n$, and mean density will both contribute to higher $f_\mathrm{grav}$ and subsequently higher $\epsilon_\mathrm{ff}$. This model implies that the SFR is set solely by mass available above $n_\mathrm{SF}\approx10^{4.5}$ cm$^{-3}$, and that the depletion time of this dense gas mass is constant.
\par
A fixed density threshold also has several other testable predictions:
\begin{itemize}
    \item Higher $\mathcal{M}$ increases $\sigma_\mathrm{v}$ and so contributes to \textit{higher} $f_\mathrm{grav}$, shorter total gas depletion times, and ultimately higher $\Sigma_\mathrm{SFR}$. As a result, for a given $\mathcal{M}$, only the increase in gas mass ($\Sigma_\mathrm{mol}$) is important for increasing $\Sigma_\mathrm{SFR}$.
    \item Higher mean densities correlate with \textit{higher} $f_\mathrm{grav}$, shorter total gas depletion times, and ultimately higher $\Sigma_\mathrm{SFR}$.
\end{itemize}

The \citet{Burkhart:2018} and \citet{Padoan:2011} models predict that $n_\mathrm{SF}$ varies, but these formalisms have different interpretations as to why this variation occurs. \citet{Burkhart:2018} argue that variations in $\epsilon_\mathrm{ff}$ may largely be linked to evolutionary changes in the $n-$PDF shape such that the $n-$PDF evolves from a lognormal shape to a composite lognormal and powerlaw over time, and the PL tail develops a shallower slope as more gas becomes gravitationally-bound (cf. \citealt{Ballesteros-Paredes:2011a}). 
\par
There is a dependence also on turbulence similar to that of the \citet{Padoan:2011} model, but \citet{Burkhart:2018} argue that this dependence becomes less important as clouds become more evolved (i.e. $\alpha_\mathrm{PL}$ becomes more shallow). The \citet{Burkhart:2018} model also does not depend on explicitly on $\alpha_\mathrm{vir}$, whereas virialization of the gas is included in the \citet{Padoan:2011} definition of $n_\mathrm{SF}$. 
\par
In the \citet{Burkhart:2018} and \citet{Padoan:2011} models, higher $\mathcal{M}$ has the effect of increasing $n_\mathrm{SF}$. Based on this, the following qualitative predictions can then be made about both of these models:
\begin{itemize}
    \item Clouds with higher $\mathcal{M}$ (higher $P_\mathrm{turb}$) can have a broader $n-$PDF and higher $n_\mathrm{SF}$, \textit{lower} $f_\mathrm{grav}$, \textit{lower} star formation efficiencies, and ultimately smaller $\Sigma_\mathrm{SFR}$.
    \item Higher temperatures reduce $\mathcal{M}$, and may \textit{decrease}  $n_\mathrm{SF}$ and potentially \textit{increase} $f_\mathrm{grav}$ and \textit{enhance} star formation efficiencies. This effect will likely be much smaller than changes in $P_\mathrm{turb}$, which can span $>5$ orders of magnitude, compared to $T_\mathrm{kin}$ which only can span $\sim2$ orders of magnitude.
\end{itemize}
\par
We note that the major difference between the predictions for a fixed  $n_\mathrm{SF}$ and one that varies as $n_\mathrm{SF}\propto \mathcal{M}^2$ is whether $\mathcal{M}$ directly enhances or suppresses star formation. For the LN+PL B18 models, smaller values of $\alpha_\mathrm{PL}$ (i.e. shallower slopes) will result in more mass in the PL tail and higher $f_\mathrm{grav}$, which will ultimately \textit{enhance} $\epsilon_\mathrm{ff}$. We consider these differences when we are comparing the predictions of these analytical models with our data in the \S \ref{sec:results}.

\subsection{Connecting Observations to Theory} \label{sec:model_grid}

\begin{figure}[tb]
    \centering
    \includegraphics[width=0.49\textwidth]{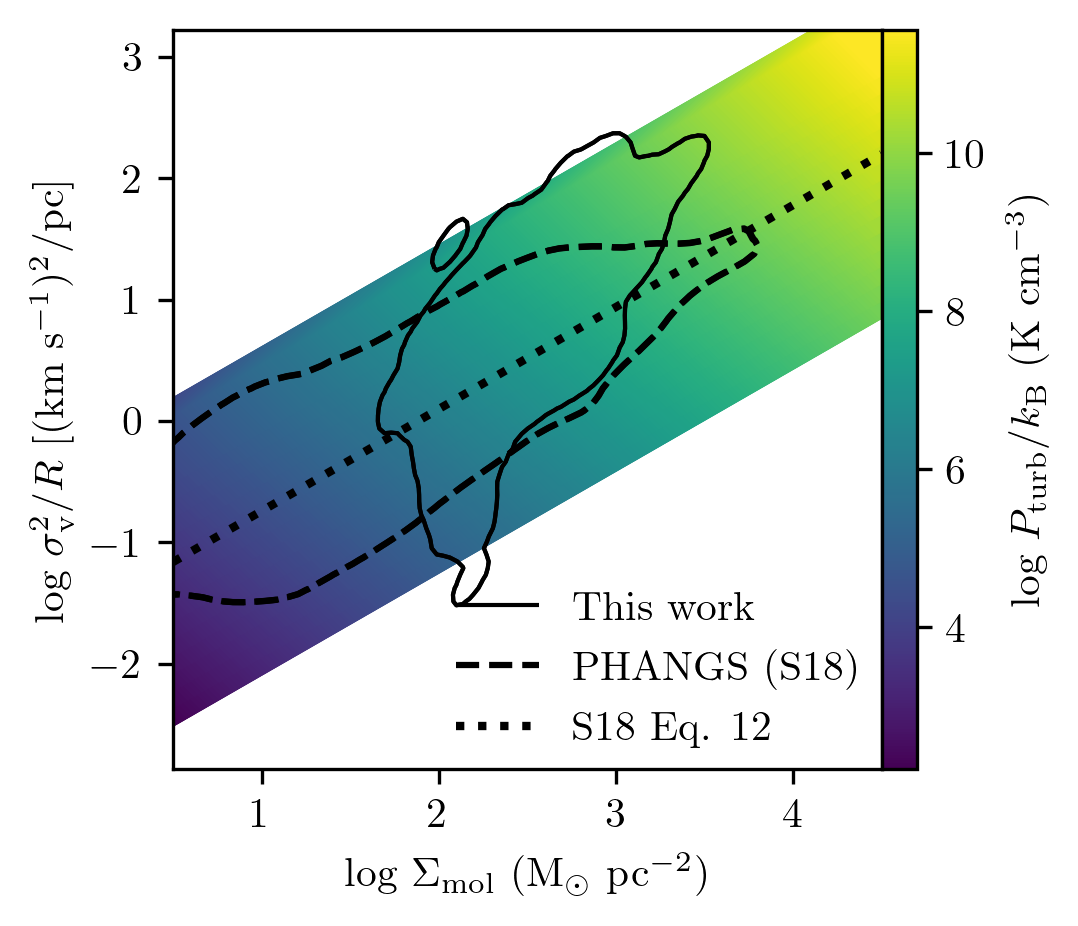}
    \caption{The grid of cloud coefficients, $\sigma_\mathrm{v}/R^2$, (cf. \citealt{Heyer:2009}) and molecular gas surface densities, $\Sigma_\mathrm{mol}$, used as inputs to the lognormal-only analytical models, colored by turbulent pressure, $P_\mathrm{turb}$. The $x-$axis is proportional to model mean density. The data are shown as the contour density plots, with our sample contained within the solid black line, and the locus of the  \citet{Sun:2018} cloud-scale measurements as the dashed black contour. We use Eq. 12 from \citet{Sun:2018} to define the mean relationship between $\Sigma_\mathrm{mol}$ and $\sigma_\mathrm{v}$ in our model parameter space (dotted black line).}
    \label{fig:grid}
\end{figure}

To compare models with data from our sample, we create a sets of models for each of the three, unique prescriptions. We produce LN models over the $\sigma_\mathrm{v}^2/R-\Sigma_\mathrm{mol}$ parameter space that encompasses our data in Fig. \ref{fig:grid}. For comparison, we also show the \citet{Sun:2018} measurements of cloud-scale observations in nearby galaxies in Fig. \ref{fig:grid}. The mean trend of this parameter space is the empirical trend between velocity dispersion and molecular gas surface density found by \citet{Sun:2018} for a subset of PHANGS galaxies at cloud-scales:
\begin{equation}\label{eq:sun18}
    \mathrm{log_{10}}\ \left(\frac{\sigma_\mathrm{v}}{\mathrm{km\ s}^{-1}} \right) = \mathrm{log_{10}}\ \left(\frac{\Sigma_\mathrm{mol}}{10^2\ M_\odot\ \mathrm{pc}^{-2}} \right) + 0.85
\end{equation}
This is shown in Fig. \ref{fig:grid} as the dotted line.  \citet{Sun:2018} assume a radius of 40 pc for their sources. Since we assume a fixed LOS depth of 100 pc, we scale the right side of Eq. \ref{eq:sun18} by a factor of $\sqrt{100\ \mathrm{pc}/40\ \mathrm{pc}}$.
\par
The vertical spread in Fig. \ref{fig:grid} is from variations in $\sigma_\mathrm{v}$ that we consider for the LN models, and roughly reproduces the spread in $\sigma_\mathrm{v}^2/R-\Sigma_\mathrm{mol}$ seen in the \citet{Sun:2018} sample as well as our sample of galaxies. For the LN+PL B18 models, we impose Eq. \ref{eq:sun18} and consider a range of $\alpha_\mathrm{PL}=1.0-2.3$, instead of the range in $\sigma_\mathrm{v}$ considered for the LN models.
\par
We note that the observational efficiency is estimated by taking the ratio of the free-fall time to the depletion time,  $\epsilon_\mathrm{ff}=t_\mathrm{ff}/t_\mathrm{dep}$.  As \citet{Burkhart:2019} point out, a more physically-meaningful efficiency for observations of individual star forming clouds may be the local \textit{instantaneous} efficiency of star formation at time $t$, $\epsilon_\mathrm{inst}\approx\epsilon_0(t)\, f_\mathrm{grav}(t)$, where $\epsilon_\mathrm{inst}< \epsilon_\mathrm{ff}$. This in particular applies to clouds that have evolved sufficiently to have a powerlaw tail, where $f_\mathrm{grav}$ is the fraction of mass that is gravitationally-bound (and therefore in the powerlaw tail). $\epsilon_\mathrm{ff}$ itself is an \textit{average} of the star formation efficiency of a cloud over the free-fall time, which is not directly observable. For simplicity we assume our observational estimates of $\epsilon_\mathrm{ff}$ are indeed sensitive to the theoretical definition of $\epsilon_\mathrm{ff}$. We also compare observational trends of $\epsilon_\mathrm{ff}$ with model trends of $f_\mathrm{grav}\propto \epsilon_\mathrm{inst}$.
\par
We adopt a constant, local efficiency of $\epsilon_0=1\%$ for all models. This value of $\epsilon_0$ is consistent with estimates of $\epsilon_\mathrm{ff}$ from observations, and simulations imply low efficiencies, which range from $0.01-\lesssim20\%$ \citep{Evans:2009,Lada:2010,OstrikerShetty:2011,Krumholz:2014,Zamora-AvilesVazquez-Semadeni:2014,Lee:2016,Semenov:2017,Grudic:2018}. We summarize the observational measurements and the quantities that we estimate from them in the top half of Table \ref{tab:obs_vs_model}. In the bottom half of Table \ref{tab:obs_vs_model} we list the equations used to estimate the same quantities using model outputs.

\par

\begin{table*}[tb]
    \centering
    \begin{threeparttable}
    \caption{Observational quantities (top panel) compared to model inputs and outputs (bottom panel).  \label{tab:obs_vs_model}}
    \begin{tabular}{llr@{$\qquad$}lrrr}
    \toprule
    & Observational Estimate   &   Eqn.  &   Unit  \\
    \midrule
    &   $\Sigma_\mathrm{mol,obs}$   &   $\alpha_\mathrm{CO}\,(I_\mathrm{CO}\,/$ K km s$^{-1}$)  &   $M_\odot$ pc$^{-2}$   \\
    &   $\Sigma_\mathrm{mol,dense,obs}$   &   $3.2\,  \alpha_\mathrm{CO} \,(I_\mathrm{HCN}\,/$ K km s$^{-1}$)   &      $M_\odot$ pc$^{-2}$   \\
    (1) &   $\Sigma_\mathrm{SFR,obs}$   &   $1.14\times10^{-29}\,(L_{93\,\mathrm{GHz}}\,/$ erg s$^{-1}$ Hz)    &      $M_\odot$ yr$^{-1}$ kpc$^{-2}$ \\
    (2) &   $P_\mathrm{turb,obs}$   &   $(3/2)\,\Sigma_\mathrm{mol}\sigma_\mathrm{v}^2/R/k_\mathrm{B}$    &      K cm$^{-3}$  \\
    (3) &   $t_\mathrm{dep,obs}$    &   $\Sigma_\mathrm{mol}/\Sigma_\mathrm{SFR}$   &   yr  \\
    &   $n_{0,\mathrm{obs}}$     &    $\Sigma_\mathrm{mol}/R$    & kg cm$^{-3}$ \\
    &   $t_\mathrm{ff,obs}$ &   $\sqrt{3\pi/32\,G\,n_{0,\mathrm{obs}}}$    &   yr  \\
    (4) &   $\epsilon_\mathrm{ff,obs}$  &   $t_\mathrm{ff,obs}/t_\mathrm{dep,obs}$  &   -- \\
    (5) &   $f_\mathrm{dense}$  &   $3.2\,I_\mathrm{HCN}/I_\mathrm{CO}$ &   --  \\ 
    (6) &   $t_\mathrm{dep,dense,obs}$    &   $\Sigma_\mathrm{mol,dense}/\Sigma_\mathrm{SFR}$   &   yr  \\
    \midrule
    &   Model Estimates   &   Eqn.    &   Unit \\
    \midrule
    (1) &   $\Sigma_\mathrm{SFR,model}$   &       $\epsilon_\mathrm{ff,model}\times\, \Sigma_\mathrm{mol,grid}/t_\mathrm{ff,grid}$  &  $M_\odot$ yr$^{-1}$ kpc$^{-2}$   \\
    (2) &   $P_\mathrm{turb,model}$   &   $(3/2)\,n_0 \sigma_\mathrm{v}^2/k_\mathrm{B}$    &      K cm$^{-3}$ \\
    (3) &   $t_\mathrm{dep,model}$  &   $1/\epsilon_\mathrm{ff,model}\times\, t_\mathrm{ff,grid}$    &   yr \\
    (4) &   $\epsilon_\mathrm{ff,model}$    & Eq. (13) or (27) in \citet{Burkhart:2018}   &   --  \\
    (5) &   $f_\mathrm{grav}$   & Eq. \ref{eq:fgrav}, Eq. (20) in \citet{Burkhart:2019}   &   -- \\
    (6) &   $t_\mathrm{dep,grav}$  &   $f_\mathrm{grav}/\epsilon_\mathrm{ff,model}\times\, t_\mathrm{ff,grid}$  &   yr  \\
    (6) &   $t_\mathrm{dep,dense,model}$  &   $f({n>10^{4.5}\,\mathrm{cm}^{-3}})/\epsilon_\mathrm{ff,model}\times\, t_\mathrm{ff,grid}$ &   yr \\
    \bottomrule
    \end{tabular}
    \begin{tablenotes}
    \item[]\textsc{Notes} -- We number matching observational and model analogs in the leftmost column. We fix $T_\mathrm{kin}=45$ K, $b=0.4$, and $\beta\rightarrow\infty$ in this analysis, and we model over the observed ranges of $\Sigma_\mathrm{mol}$ and $\sigma_\mathrm{v}$. Quantities with the subscript `grid' are input from the grid shown in Fig. \ref{fig:grid}.
    \end{tablenotes}
    \end{threeparttable}
\end{table*}

\section{Results}\label{sec:results}

In this section we review general trends between observation-based quantities and the predictions of those from analytical models of star formation. We present a series of plots showing the observational quantities listed in the top half of Table \ref{tab:obs_vs_model} that we will use to compare against the model predictions described in the bottom half of Table \ref{tab:obs_vs_model}. We focus on variations in star formation timescales (depletion time and free-fall time), star formation efficiency, and the dense gas fraction. We do not directly assess variations in emissivity in this paper, and defer an analysis of emissivity (and therefore conversion factors) to \citet{Bemis:2020} and Bemis \& Wilson, in prep., where we present modelling of the HCN and CO emissivities for our sample of galaxies on a pixel-by-pixel basis.
\par
The results from analytical models are compared to the data in Figures \ref{fig:KS_data}-\ref{fig:tdep_dense}. We consider fiducial models which correspond to the parameters that best fit the observed Kennicutt-Schmidt relationship (see Fig. \ref{fig:KS_plots}). For each model, we calculate the $\chi^2$ of the data relative to this fiducial model, and this value is shown in each panel. We note that the best fit value for $\alpha_\mathrm{PL}$ is 1.9 for the LN+PL models. The best fit-relationship between $\Sigma_\mathrm{mol}-\sigma_v$ for the LN-only models is in agreement with the \citet{Sun:2018} relationship (Eq. \ref{eq:sun18}), within uncertainty. 
\par
We show separate plots with only the data and their measurement uncertainties, followed by plots comparing the data with the models. We refer to Spearman rank coefficients as a measure of the strength and direction of correlations between two parameters. For reference, a coefficient of $r_s=0$ indicates that there is no monotonic relationship between the two parameters, negative values indicate negative monotonic relationships, and positive values indicate positive monotonic relationships. In the following discussion, we use the following definitions:  $0.1\le|r_s|<0.4$ is considered a weak correlation, $0.4\le|r_s|<0.7$ is considered a moderate correlation, and $|r_s|\ge0.7$ is considered a strong correlation.

\subsection{The Kennicutt-Schmidt Relationship}

\begin{figure}[tb]
    \centering
    \includegraphics[width=0.45\textwidth]{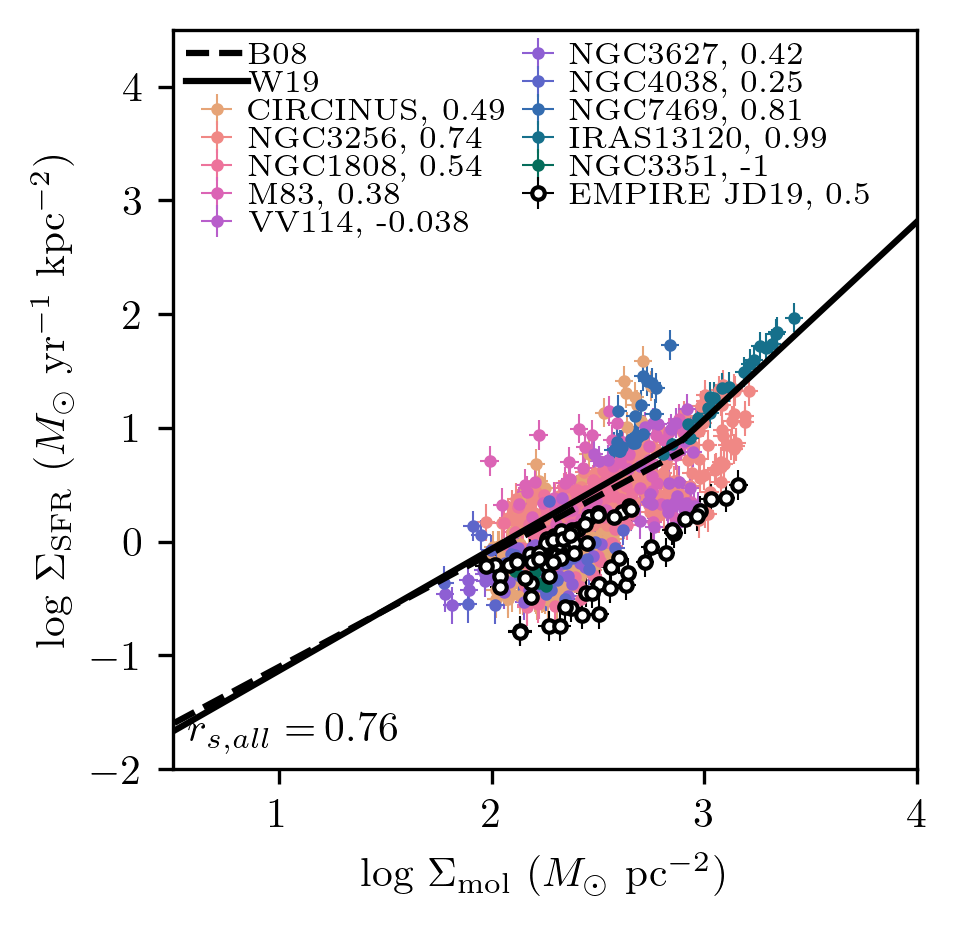}
    \caption{The Kennicutt-Schmidt relationship for our galaxies and the EMPIRE sample. Data is shown for our sample (colorized by galaxy) and the EMPIRE galaxies (white points). We also include fits to this relationship from previous studies (e.g. nearby disk galaxies from \citealt{Bigiel:2008}, B08, and (U)LIRGs from \citet{Wilson:2019}, W19). Spearman rank coefficients are shown for individual galaxies next to their name in the legend, and the coefficient for the combined sample is shown in the lower left corner. On average, the galaxies in the EMPIRE sample have lower $\Sigma_\mathrm{SFR}$ than the galaxies in our sample, despite similar gas surface densities. Compared to the normal disk galaxies of the EMPIRE sample, the galaxies in our sample are more extreme and include mergers, U/LIRGs, and galaxy centers (cf. Table \ref{tab:source_data_props}.)\label{fig:KS_data}}
\end{figure}

\begin{figure*}[tb]
    \centering
    \includegraphics[width=0.95\textwidth]{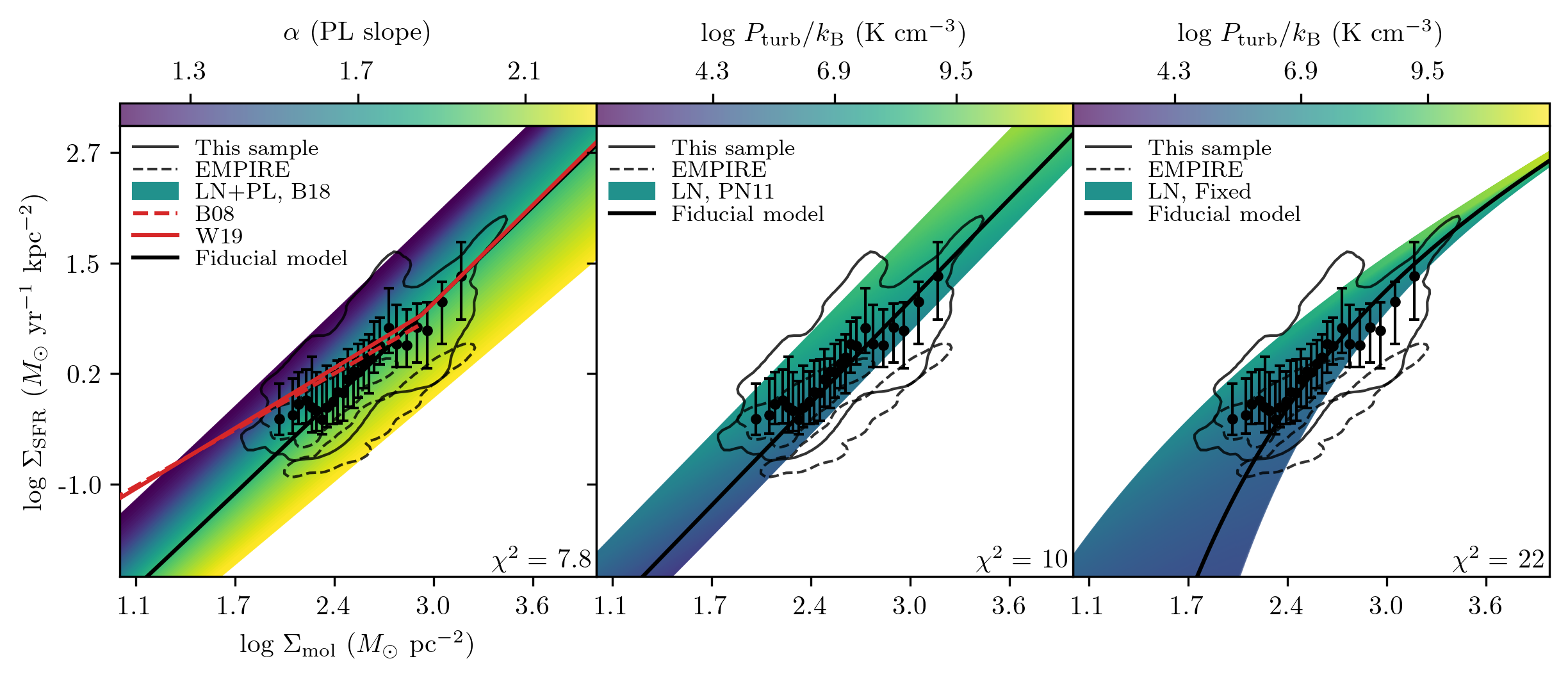}
    \caption{The Kennicutt-Schmidt relationship for our galaxies (solid black contours) and the EMPIRE sample (dashed black contours), compared to the model predictions. We also include fits to this relationship from previous studies (see Fig. \ref{fig:KS_data} caption). The three models we consider in this analysis are: the LN+PL B18 model with varying $n_\mathrm{SF}$ (left), the LN model with the \citet{Padoan:2011} $n_\mathrm{SF}$ (center), and the LN model with a fixed $n_\mathrm{SF}=10^{4.5}$ cm$^{-3}$ (right). The LN+PL B18 models are colored by $\alpha_\mathrm{PL}$, and the LN models are colored by $P_\mathrm{turb}$. $\chi^2$ relative to the fiducial model is shown in the lower right corner of each plot. The median values of the data are shown with the black datapoints, and the errorbars correspond to the $1-\sigma$ spread in each bin.} Each model is able to reproduce some of the scatter of the Kennicutt-Schmidt relationship when considering variations in $\alpha_\mathrm{PL}$ (LN+PL B18 models) and $P_\mathrm{turb}\propto\sigma_\mathrm{v}^2$ (LN models) for a given gas surface density. \label{fig:KS_plots}
\end{figure*}
We begin by presenting the Kennicutt-Schmidt (KS) relationship of our data in Fig. \ref{fig:KS_data}. In Fig. \ref{fig:KS_plots} we compare the data to model predictions of $\Sigma_\mathrm{SFR}$, which are determined using the relevant equation in Table \ref{tab:obs_vs_model} and the parameter spaces discussed in \S \ref{sec:model_grid}. 

\begin{enumerate}
    \item \textit{LN PN11:} The LN models with the varying \citet{Padoan:2011} threshold produce a steeper relationship than observed at lower gas surface densities. The (U)LIRGs studied in \citet{Wilson:2019} show a double power-law Kennicutt-Schmidt relationship, with lower $\Sigma_\mathrm{mol}$ galaxies returning a slope close to unity, similar to the results of \citet{Bigiel:2008}. The higher-$\Sigma_\mathrm{mol}$ galaxies instead have a steeper Kennicutt-Schmidt slope of 1.74. For the parameter space we consider, the trend of the Kennicutt-Schmidt relationship predicted by the PN11 models is in agreement with a slope$>1$.
    \item \textit{LN+PL B18:} Similar to the PN11 models,  the LN+PL B18 models produce a steeper relationship than observed at lower gas surface densities, and match the steep slope observed by \citet{Wilson:2019} for U/LIRGs. We find larger $\alpha_\mathrm{PL}$ results in lower $\Sigma_\mathrm{SFR}$ (left panel of Fig. \ref{fig:KS_plots}), in agreement with the predictions from \citet{Burkhart:2018}. \citet{Burkhart:2018} argue that larger $\alpha_\mathrm{PL}$ results in lower $\epsilon_\mathrm{ff}$  (and higher $\Sigma_\mathrm{SFR}$), which is seen for some Milky Way clouds.
    \item \textit{LN Fixed:} These fixed-threshold models produce a trend in $\Sigma_\mathrm{SFR}$ with $\Sigma_\mathrm{mol}$ that is not seen in the data. Lower $\Sigma_\mathrm{mol}$ values produce a steeper Kennicutt-Schmidt relation, while higher $\Sigma_\mathrm{mol}$ produce a shallower Kennicutt-Schmidt relation, in conflict with what is observed.
\end{enumerate}

Figure \ref{fig:KS_plots} demonstrates that each model is able to reproduce some of the scatter of the Kennicutt-Schmidt relationship when considering variations in $\alpha_\mathrm{PL}$ (LN+PL B18 models) and $\sigma_\mathrm{v}$ (LN models) for a given gas surface density. No models produce a multi-slope Kennicutt-Schmidt relationship consistent with the slopes measured in previous studies (e.g. \citealt{Wilson:2019}). However, we note that changing the slope of the imposed relationship between $\sigma_\mathrm{v}$ and $\Sigma_\mathrm{mol}$ (Fig. \ref{fig:grid}) has the effect of changing the slope of the Kennicutt-Schmidt relation for all of these models. Lower slopes, e.g. log $\sigma_\mathrm{v}\propto 0.1\mathrm{log}\ \Sigma_\mathrm{mol}$, produce a shallower Kennicutt-Schmidt relation. We cannot exclude the possibility that other model parameters (i.e. $b$, $\beta$, and $T_\mathrm{kin}$) may have underlying trends with $\Sigma_\mathrm{mol}$ or $\sigma_\mathrm{v}$, which could in turn also produce a varying Kennicutt-Schmidt slope. Additionally, $\epsilon_0$ could vary with $\Sigma_\mathrm{mol}$, for example if a higher $\Sigma_\mathrm{SFR}$ results in higher stellar feedback and a lower $\epsilon_0$. Variations in PL slope may also contribute to changes in the slope of the Kennicutt-Schmidt relation. \citet{Federrath:2013} find that decreases in power law slope coincide with enhanced star formation efficiencies and vice versa \citep{Burkhart:2019,Federrath:2013}, and that this is a reflection of the increased fraction of dense gas that coincides with shallow power law slopes. To match the shallower slope of the KS relationship at gas surface densities below $\Sigma_\mathrm{mol}\approx10^{2.9}\ \mathrm{M}_\odot\ \mathrm{pc}^{-2}$ \citep[e.g.]{Bigiel:2008,Wilson:2019,Kennicutt:2021}, the average PL slope would need to increase (become steeper) towards higher gas surface densities. This would also imply lower $\epsilon_0$ towards higher $\Sigma_\mathrm{mol}$ \citep{Burkhart:2019,Federrath:2013}. Above $\Sigma_\mathrm{mol}\approx10^{2.9}\ \mathrm{M}_\odot\ \mathrm{pc}^{-2}$, the steeper KS slope found in \citet{Wilson:2019} appears consistent with a constant PL slope.

\subsection{Does HCN/CO trace the Star-forming Gas Fraction?}

\begin{figure}[tb]
    \centering
    \includegraphics[width=0.45\textwidth]{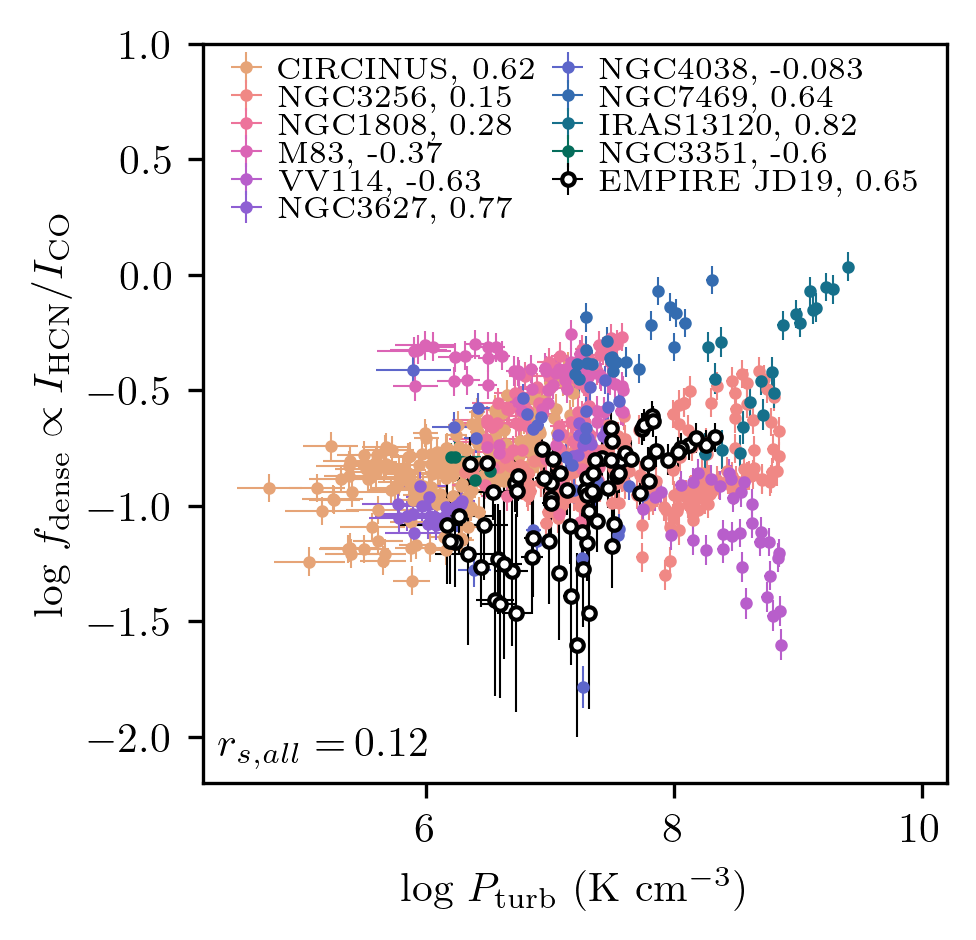}
    \caption{Dense gas fraction  as a function of turbulent pressure for our galaxies and the EMPIRE sample, where $f_\mathrm{dense}=\alpha_\mathrm{HCN}/\alpha_\mathrm{CO}\,I_\mathrm{HCN}/I_\mathrm{CO}$. Correlation coefficients are shown and data is colorized by galaxy, similar to the formatting in Fig. \ref{fig:KS_data}. The sample as a whole shows a weak, positive correlation between $f_\mathrm{dense}$ traced by $I_\mathrm{HCN}/I_\mathrm{CO}$ and $P_\mathrm{turb}$. The disk galaxies in the EMPIRE sample show a strong, positive correlation. \label{fig:fraction_vdisp_data_only}}
\end{figure}

\begin{figure*}[tb]
    \centering
    \includegraphics[width=0.95\textwidth]{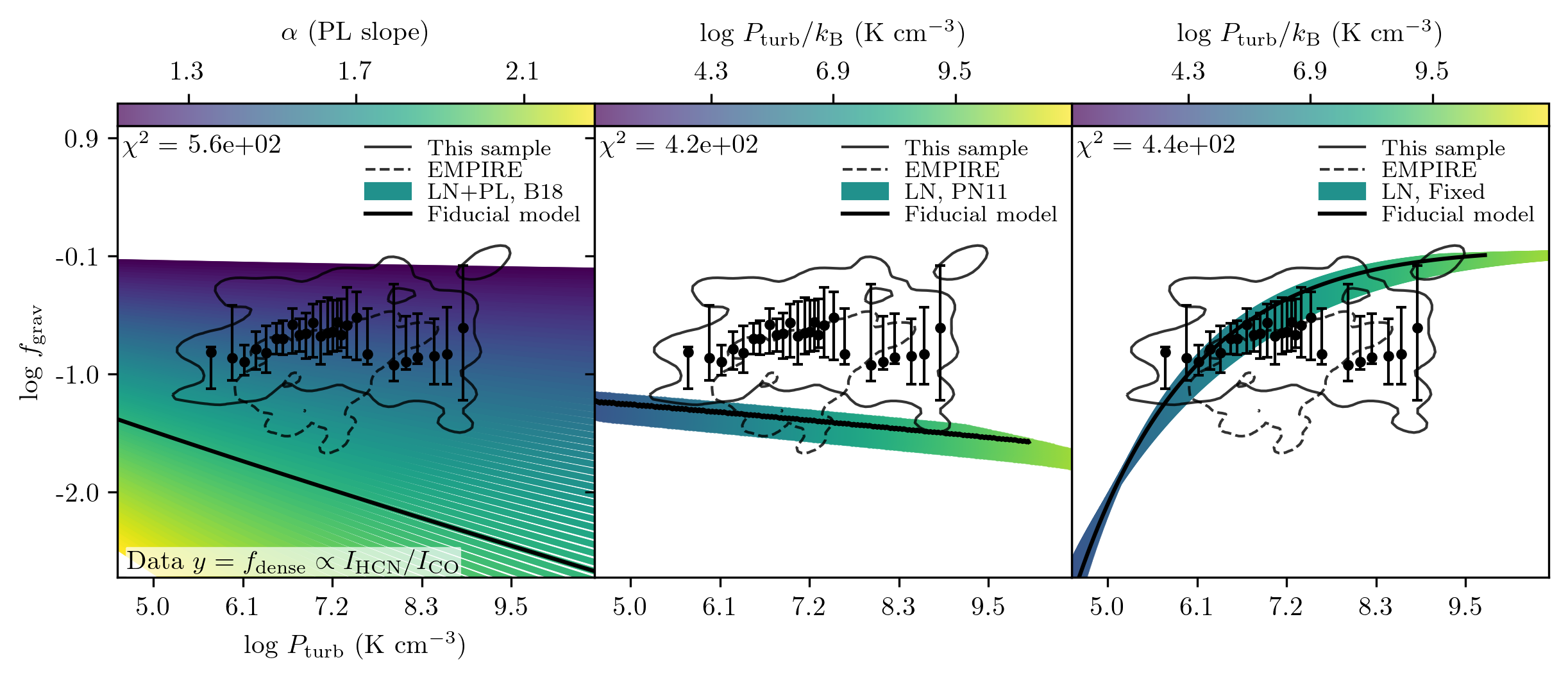}
    \includegraphics[width=0.95\textwidth]{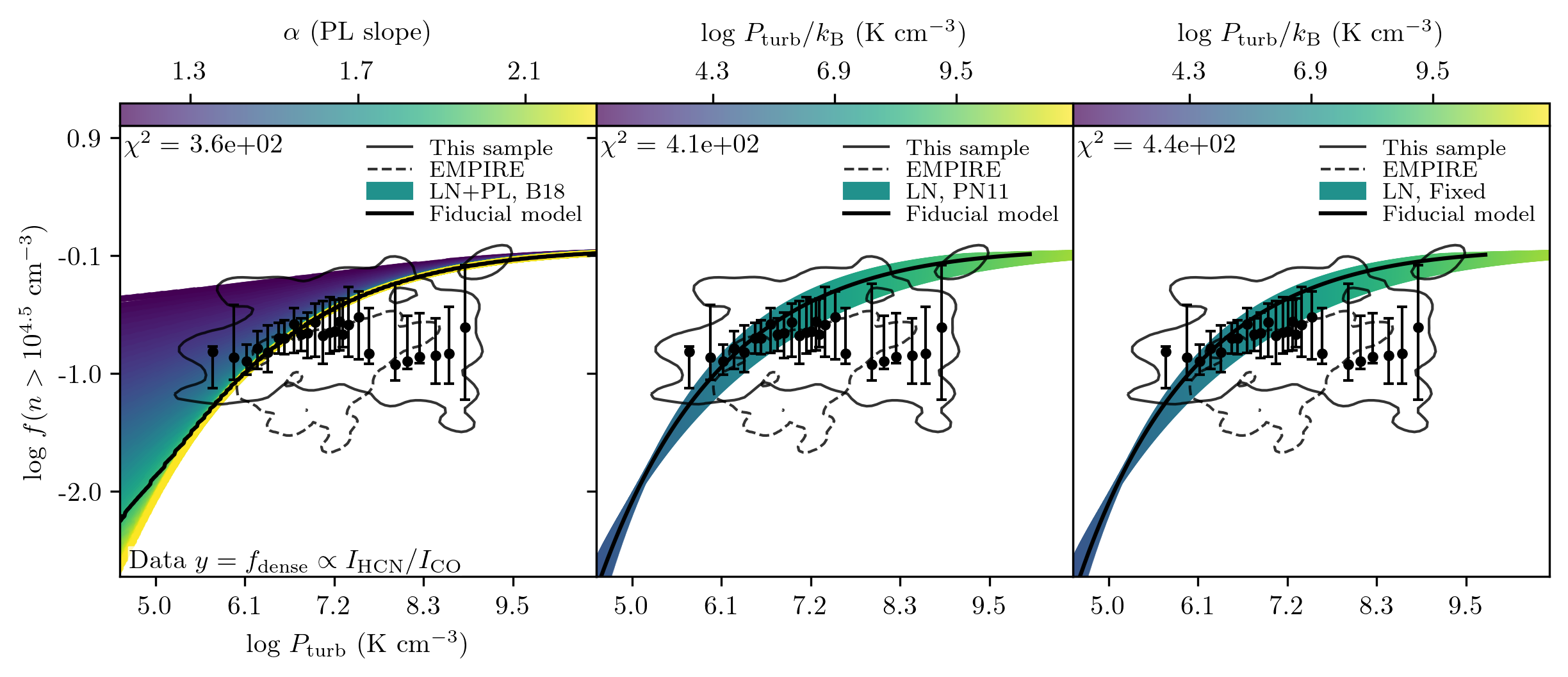}
    \caption{Model $f_\mathrm{grav}$ (top) and $f(n>10^{4.5}\ \mathrm{cm}^{-3})$ (bottom) as a function of $P_\mathrm{turb}$, compared to the data from our sample (solid black contours) and the EMPIRE sample (dashed black contours). Note that $f_\mathrm{grav}\equiv f_\mathrm{dense}$ for models with $n_\mathrm{SF}=10^{4.5}$ cm$^{-3}$. The LN+PL B18 models are colored by $\alpha_\mathrm{PL}$, and the LN models are colored by $P_\mathrm{turb}$. $\chi^2$ relative to the fiducial model is shown in the upper left corner of each plot. The median values of the data are shown with the black datapoints, and the errorbars correspond to the $1-\sigma$ spread in each bin.} Model $f_\mathrm{grav}$ decreases with increasing $P_\mathrm{turb}$ for the models with varying thresholds (i.e. LN+PL B18 and LN P11, left two columns). $f_\mathrm{grav}$ predicted by the fixed-density threshold models is identical to $f(n>10^{4.5}$ cm$^{-3}$) (right column) and increases with $P_\mathrm{turb}$. Model $f(n>10^{4.5}$ cm$^{-3})$ increases with $P_\mathrm{turb}$ for all models. From this result, the observed positive trend in $I_\mathrm{HCN}/I_\mathrm{CO}$ with $P_\mathrm{turb}$ is more consistent with $I_\mathrm{HCN}/I_\mathrm{CO}$ being a better tracer of $f(n>10^{4.5}$ cm$^{-3})$ than $f_\mathrm{grav}$.  \label{fig:gas_fractions}
\end{figure*}

We compare the observed dense gas fraction ($f_\mathrm{dense}=\alpha_\mathrm{HCN}/\alpha_\mathrm{CO}\,I_\mathrm{HCN}/I_\mathrm{CO}$) with estimates of turbulent pressure, $P_\mathrm{turb}$, in Fig. \ref{fig:fraction_vdisp_data_only}. Our data, on average, show a weak positive trend between $f_\mathrm{dense}$ and $P_\mathrm{turb}$, and this relation also holds within most of the individual galaxies in our sample. We plot model predictions of $f_\mathrm{grav}$ vs. $P_\mathrm{turb}$ in Fig. \ref{fig:gas_fractions}, and show the outline of the data relationship between $f_\mathrm{dense}$ and $P_\mathrm{turb}$ in the background. For comparison, in the bottom rows of Fig. \ref{fig:gas_fractions} we plot the fraction of gas above a fixed density $n=10^{4.5}$ cm$^{-3}$ for all of the three model $n-$PDFs prescriptions.

\begin{enumerate}
    \item \textit{LN PN11:} The LN varying threshold models also predict a negative trend between $f_\mathrm{grav}$ and $P_\mathrm{turb}$. On average $f_\mathrm{grav}$ is lower than $f_\mathrm{dense}$ predicted by the data. Similar to the LN+PL B18 models, we see a positive trend between $f(n>10^{4.5}$ cm$^{-3}$) and $P_\mathrm{turb}$. However, there is very little spread in the model $f(n>10^{4.5}$ cm$^{-3}$) vs $P_\mathrm{turb}$ relationship. This indicates that the spread observed in the $f_\mathrm{dense}$ vs. $P_\mathrm{turb}$ relationship in our data is not well-reproduced by variations in $\sigma_\mathrm{v}$ alone.
    \item \textit{LN+PL B18:} For constant $\alpha_\mathrm{PL}$, these models predict a negative trend between $f_\mathrm{grav}$ and $P_\mathrm{turb}$, and this becomes steeper for larger $\alpha_\mathrm{PL}$. The data primarily overlap with $f_\mathrm{grav}$ for models with shallower values of $\alpha_\mathrm{PL}$. In contrast, we see a positive trend between $f(n>10^{4.5}$ cm$^{-3}$) predicted by the models and $P_\mathrm{turb}$. Lower values of $\alpha_\mathrm{PL}$ produce a flatter relationship between $f(n>10^{4.5}$ cm$^{-3}$) at low $P_\mathrm{turb}$. This results in a broader spread in this relationship at lower $P_\mathrm{turb}$. In general, the spread observed in the $f_\mathrm{dense}$ vs. $P_\mathrm{turb}$ relationship in our data is not well-reproduced by variations in $\alpha_\mathrm{PL}$ alone.  
    \item \textit{LN Fixed:} $f_\mathrm{grav}$ predicted by the fixed-density threshold models is identical to $f(n>10^{4.5}$ cm$^{-3}$), since we set the threshold for these models to $n=10^{4.5}$ cm$^{-3}$). As with the other models, we see a positive relationship between $f(n>10^{4.5}$ cm$^{-3}$) and $P_\mathrm{turb}$. The spread in the data is not well-reproduced by these models.
\end{enumerate}

The predictions of $f_\mathrm{grav}$ from the varying threshold models show negative trends with $P_\mathrm{turb}$ for constant $\alpha_\mathrm{PL}$ (LN+PL B18 models), in contrast to the positive trends observed with $f(n>10^{4.5}\ \mathrm{cm}^{-3})$ for all models. $f_\mathrm{grav}$ is equivalent to $f(n>10^{4.5}$ cm$^{-3}$) in the fixed-threshold models. From these results, the weak positive trend of $f_\mathrm{dense}\propto I_\mathrm{HCN}/I_\mathrm{CO}$ with $P_\mathrm{turb}$ seen in our data is more consistent with the positive trend between $f(n>10^{4.5}\ \mathrm{cm}^{-3})$ and $P_\mathrm{turb}$ seen in the models, and supports the conclusion that $I_\mathrm{HCN}/I_\mathrm{CO}$ is tracing gas above a roughly constant density but not necessarily the $f_\mathrm{grav}$ predicted by gravoturbulent models of star formation.

\subsubsection{The Star Formation Efficiency per Free-fall Time}

\begin{figure*}[tb]
    \centering
    \includegraphics[width=0.45\textwidth]{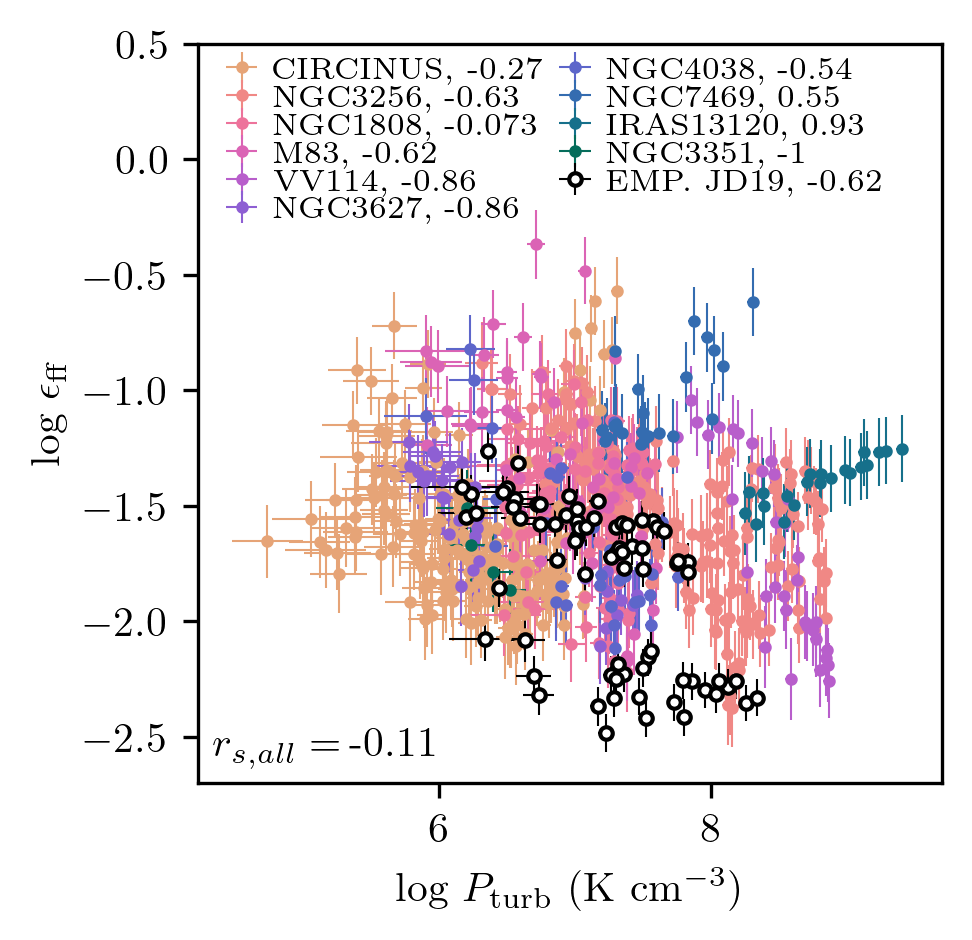}
    \includegraphics[width=0.45\textwidth]{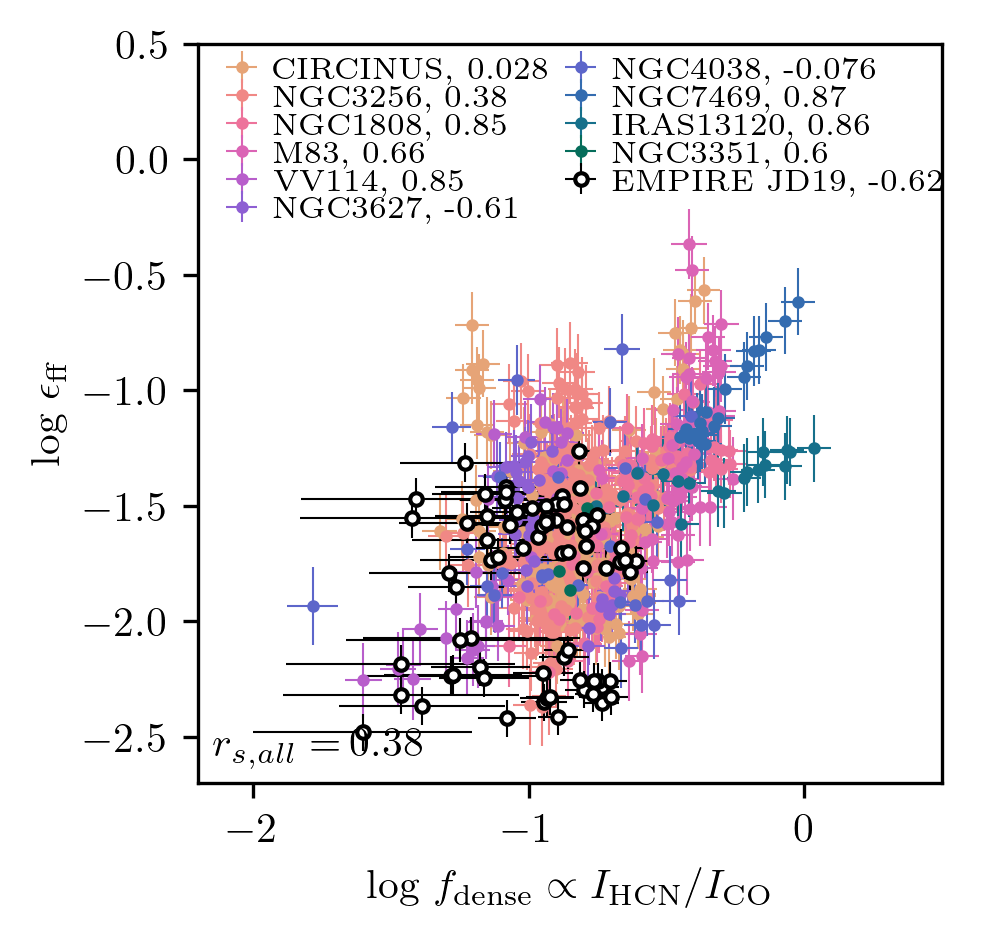}
    \caption{\textit{Left:} The efficiency per free-fall time as a function of $P_\mathrm{turb}$ for our galaxies and the EMPIRE sample. Correlation coefficients are shown and data is colorized by galaxy, similar to the formatting in Fig. \ref{fig:KS_data}. The sample as a whole shows a weak, negative correlation between $\epsilon_\mathrm{ff}$ and $P_\mathrm{turb}$ The disk galaxies in the EMPIRE sample show a moderate, negative correlation. \textit{Right:} The efficiency per free-fall time as a function of $f_\mathrm{dense}\propto I_\mathrm{HCN}/I_\mathrm{CO}$. The sample as a whole shows a moderate, positive correlation between $\epsilon_\mathrm{ff}$ and $f_\mathrm{dense}$. The EMPIRE sample shows a moderate, negative correlation. \label{fig:eff_vdisp_data}}
\end{figure*}

\begin{figure*}[tb]
    \centering
    \includegraphics[width=0.9\textwidth]{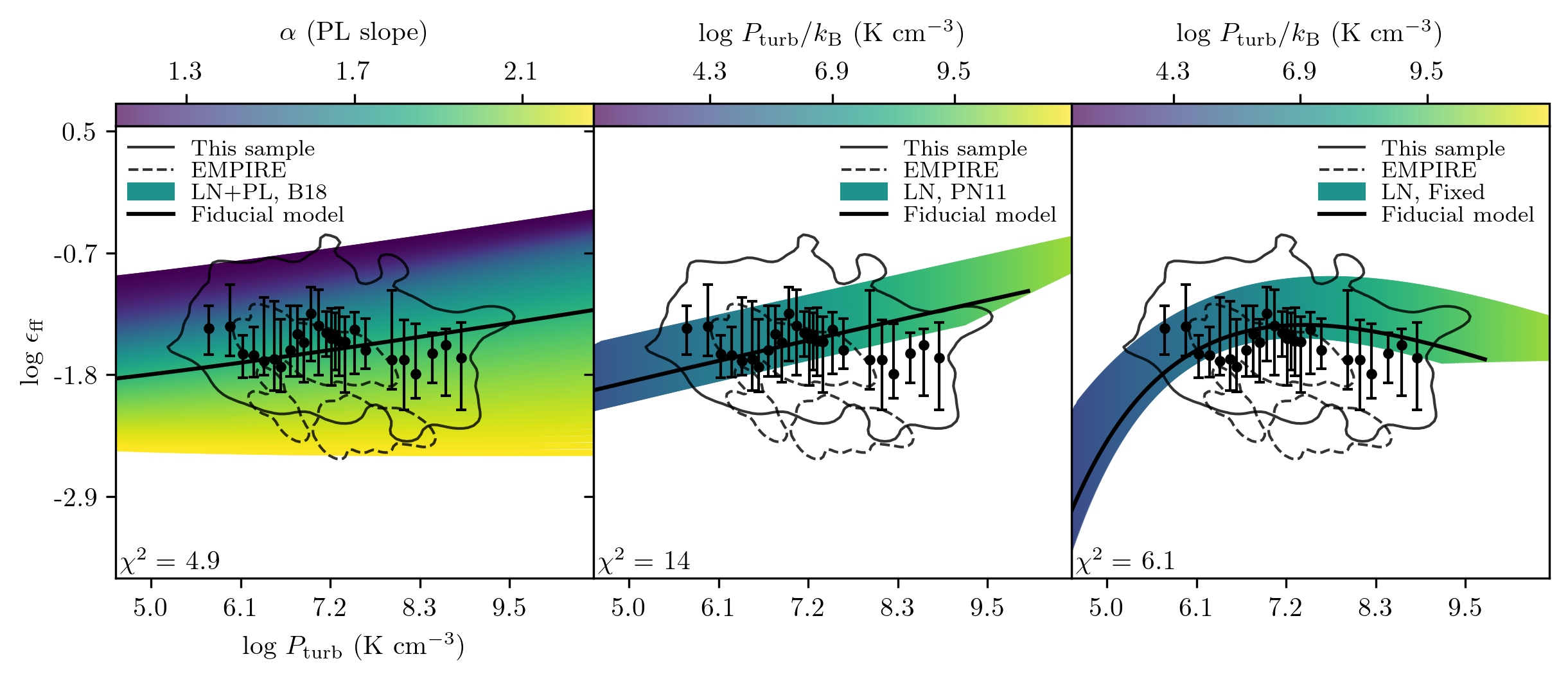}
    \includegraphics[width=0.9\textwidth]{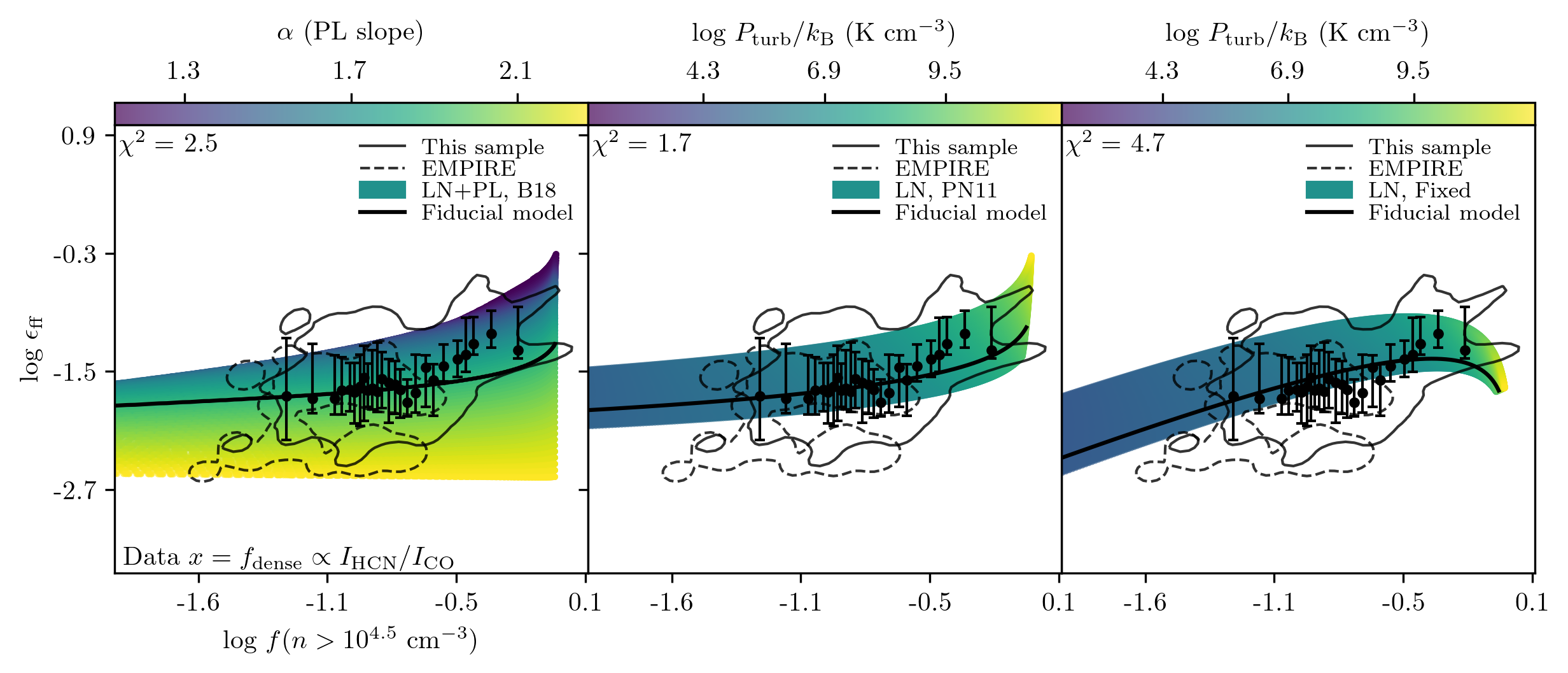}
    \includegraphics[width=0.9\textwidth]{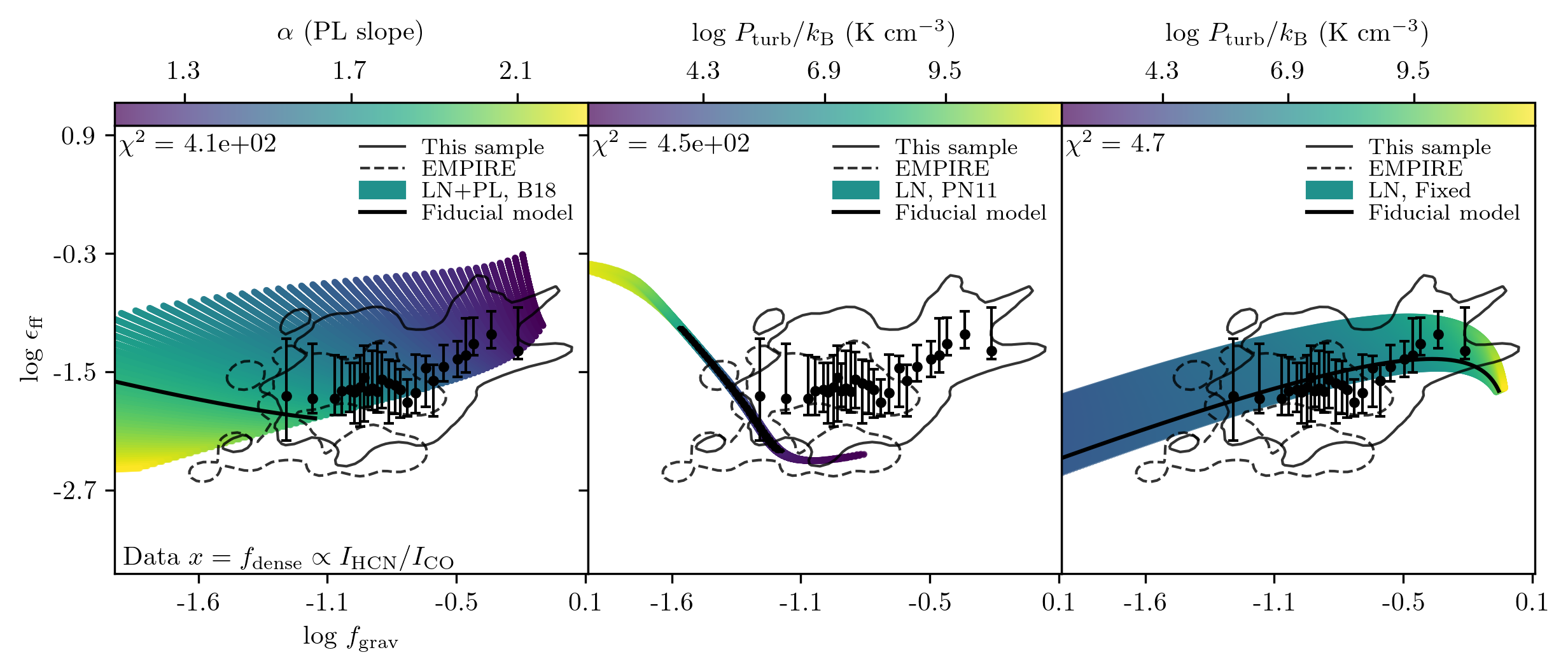}
    \caption{Model $\epsilon_\mathrm{ff}$ as a function of $P_\mathrm{turb}$ (top), $f(n>10^{4.5}$ cm$^{-3}$) (middle), and $f_\mathrm{grav}$ (bottom) compared to data from our sample (solid black contours) and the EMPIRE sample (dashed black contours). See Table \ref{tab:obs_vs_model} for information on how these values are calculated. The LN+PL B18 models are colored by $\alpha_\mathrm{PL}$, and the LN models are colored by $P_\mathrm{turb}$. $\chi^2$ relative to the fiducial model is shown in the bottom left corner of the plots in the top row, and the upper left corner of the plots in the bottom two rows. The median values of the data are shown with the black datapoints, and the errorbars correspond to the $1-\sigma$ spread in each bin. The bottom two rows show the importance of distinguishing $f_\mathrm{grav}$ from $f(n>10^{4.5}$ cm$^{-3})$ when using $I_\mathrm{HCN}/I_\mathrm{CO}$ as a tracer of the dense, star forming gas. All models considered agree better with observations when using $f(n>10^{4.5}$ cm$^{-3})$ as a proxy for the dense gas fraction traced by $I_\mathrm{HCN}/I_\mathrm{CO}$. Please see the text for a full analysis of the plots. \label{fig:eff_vdisp}
    }
\end{figure*}

As a check on the results above, we also consider $\epsilon_\mathrm{ff}$. In Fig. \ref{fig:eff_vdisp_data}, we plot our data as a function of $P_\mathrm{turb}$ (left panel) and $f_\mathrm{dense}$ (right panel). In Fig. \ref{fig:eff_vdisp} we compare the model predictions of $\epsilon_\mathrm{ff}$ with $P_\mathrm{turb}$ (top row), $f(n>10^{4.5}$ cm$^{-3}$) (middle row), and $f_\mathrm{grav}$ (bottom row). $\epsilon_\mathrm{ff}$ shows a weak negative correlation with $P_\mathrm{turb}$ and a weak positive correlation with $f_\mathrm{dense}$ (Fig. \ref{fig:eff_vdisp_data}) when considering our sample as a whole. These results are in agreement with the qualitative predictions of the varying density-threshold models  (discussed in \S \ref{sec:fgrav_theory}), where higher $\mathcal{M}\sim\sigma_\mathrm{v}$ (and therefore $P_\mathrm{turb}\sim \sigma_v^2$) yield a lower $\epsilon_\mathrm{ff}$. 

\begin{enumerate}
    \item \textit{LN PN11:} The negative trend between $\epsilon_\mathrm{ff}$ and $P_\mathrm{turb}$ seen in the data is not reproduced by the LN PN11 models (middle column, top row of Fig \ref{fig:eff_vdisp}). However, the LN PN11 models do reproduce the positive between $\epsilon_\mathrm{ff}$ and $f_\mathrm{dense}$ seen in the data when we consider $f(n>10^{4.5}\ \mathrm{cm}^{-3})$ (middle row of Fig \ref{fig:eff_vdisp}). There is some overlap between $\epsilon_\mathrm{ff}$ and model $f_\mathrm{grav}$ (bottom row of Fig \ref{fig:eff_vdisp}), but this trend does not track the positive relationship between $\epsilon_\mathrm{ff}$ and $f_\mathrm{dense}$ seen in the data.  
    \item \textit{LN+PL B18:} Similar to the LN PN11 models, the negative trend between $\epsilon_\mathrm{ff}$ and $P_\mathrm{turb}$ seen in the data is not reproduced by any single value of $\alpha_\mathrm{PL}$ for these  models (left column, top row of Fig \ref{fig:eff_vdisp}). Rather, the data overlap with these models for a range of $\alpha_\mathrm{PL}$, with higher values of $\epsilon_\mathrm{ff}$ corresponding to lower values of $\alpha_\mathrm{PL}$ and lower $P_\mathrm{turb}$.
    \par
    Also similar to the LN PN11 models, the LN+PL B18 models do reproduce the positive trend between $\epsilon_\mathrm{ff}$ and $f_\mathrm{dense}$ seen in the data when we consider $f(n>10^{4.5}\ \mathrm{cm}^{-3})$ (middle row of Fig \ref{fig:eff_vdisp}). In this case, the vertical spread in the data corresponds to variations in $\alpha_\mathrm{PL}$. Models with lower $\alpha_\mathrm{PL}$ also show an upturn in $\epsilon_\mathrm{ff}$ at higher $f(n>10^{4.5}\ \mathrm{cm}^{-3})$ consistent with the trend seen in the data.
    \par
    At first look, the positive trend between $\epsilon_\mathrm{ff}$ and $f_\mathrm{dense}$ seen in the data also appears consistent with the trend between $\epsilon_\mathrm{ff}$ and model $f_\mathrm{grav}$ (bottom row of Fig \ref{fig:eff_vdisp}). However, this requires that $f_\mathrm{dense}\propto f_\mathrm{grav}$, for which we find evidence to the contrary in the previous section. 
    \item \textit{LN Fixed:} The fixed-threshold models predict an increase in $\epsilon_\mathrm{ff}$ with $P_\mathrm{turb}$ up to $P_\mathrm{turb}\approx10^7$ K cm$^{-3}$, after which this relationship flattens and $\epsilon_\mathrm{ff}$ slightly turns over. This is not seen in our data. These models predict a positive relationship between $\epsilon_\mathrm{ff}$ and $f_\mathrm{grav}$ up to $\mathrm{log}\ f_\mathrm{grav}\approx-0.5$, above which $\epsilon_\mathrm{ff}$ turns over. There is no evidence for this turnover in our data.
\end{enumerate}

The varying threshold models are able to reproduce the positive trend seen between $\epsilon_\mathrm{ff}$ and $f_\mathrm{dense}$ when considering $f(n>10^{4.5}\ \mathrm{cm}^{-3})$. Some of the scatter is also reproduced by these models when considering a range in $\alpha_\mathrm{PL}$ or a range in $\sigma_\mathrm{v}$ (cf. Fig. \ref{fig:eff_vdisp}).

\subsubsection{Total Gas Depletion Times}

\begin{figure*}[tb]
    \centering
    \includegraphics[width=0.49\textwidth]{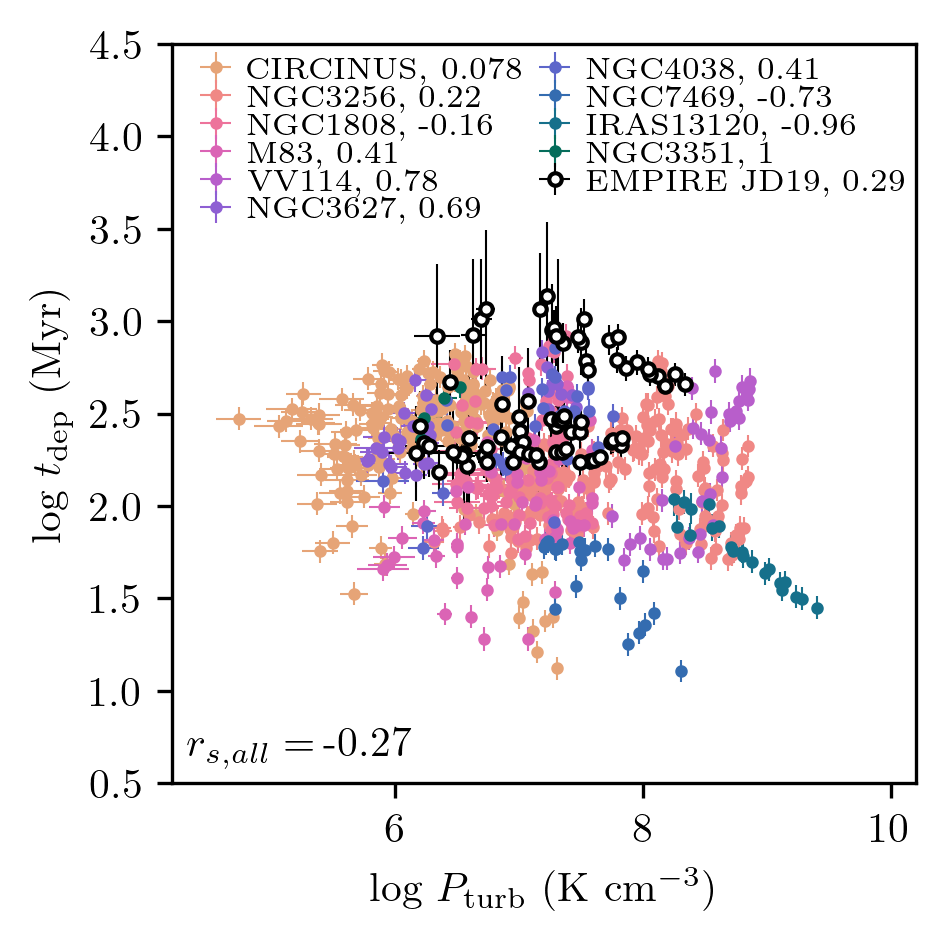}
    \includegraphics[width=0.49\textwidth]{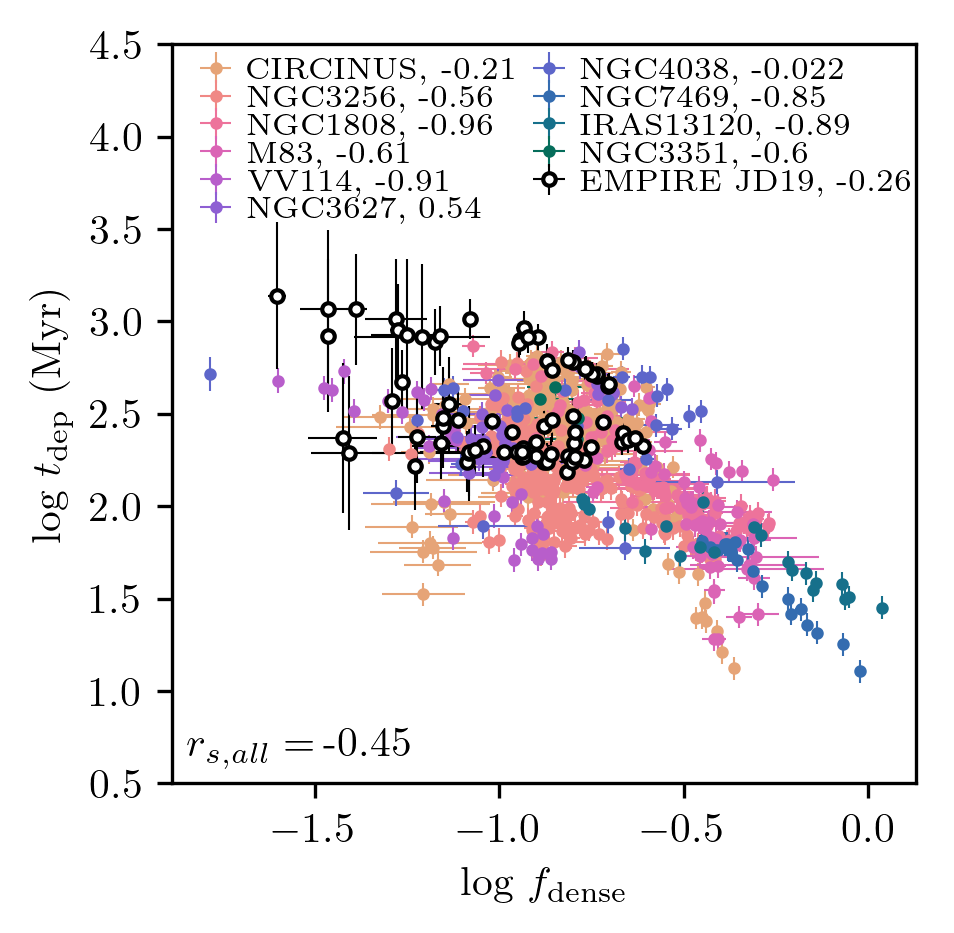}
    \caption{\textit{Left:}  Depletion time of the total molecular gas content as a function of $P_\mathrm{turb}$ for our galaxies and the EMPIRE sample. Correlation coefficients are shown and data is colorized by galaxy, similar to the formatting in Fig. \ref{fig:KS_data}. The sample as a whole shows a weak, negative correlation between $t_\mathrm{dep}$ and $P_\mathrm{turb}$. The disk galaxies in the EMPIRE sample show a weak, positive correlation. \textit{Right:} Depletion time of the total molecular gas content as a function of $f_\mathrm{dense}$ (right). The sample as a whole shows a moderate, negative correlation between $t_\mathrm{dep}$ and $P_\mathrm{turb}$. The disk galaxies in the EMPIRE sample show a weak, negative correlation. Data is formatted as it is in Fig \ref{fig:KS_data}. \label{fig:tdep_data}}
\end{figure*}

\begin{figure*}[tb]
    \centering
    \includegraphics[width=0.95\textwidth]{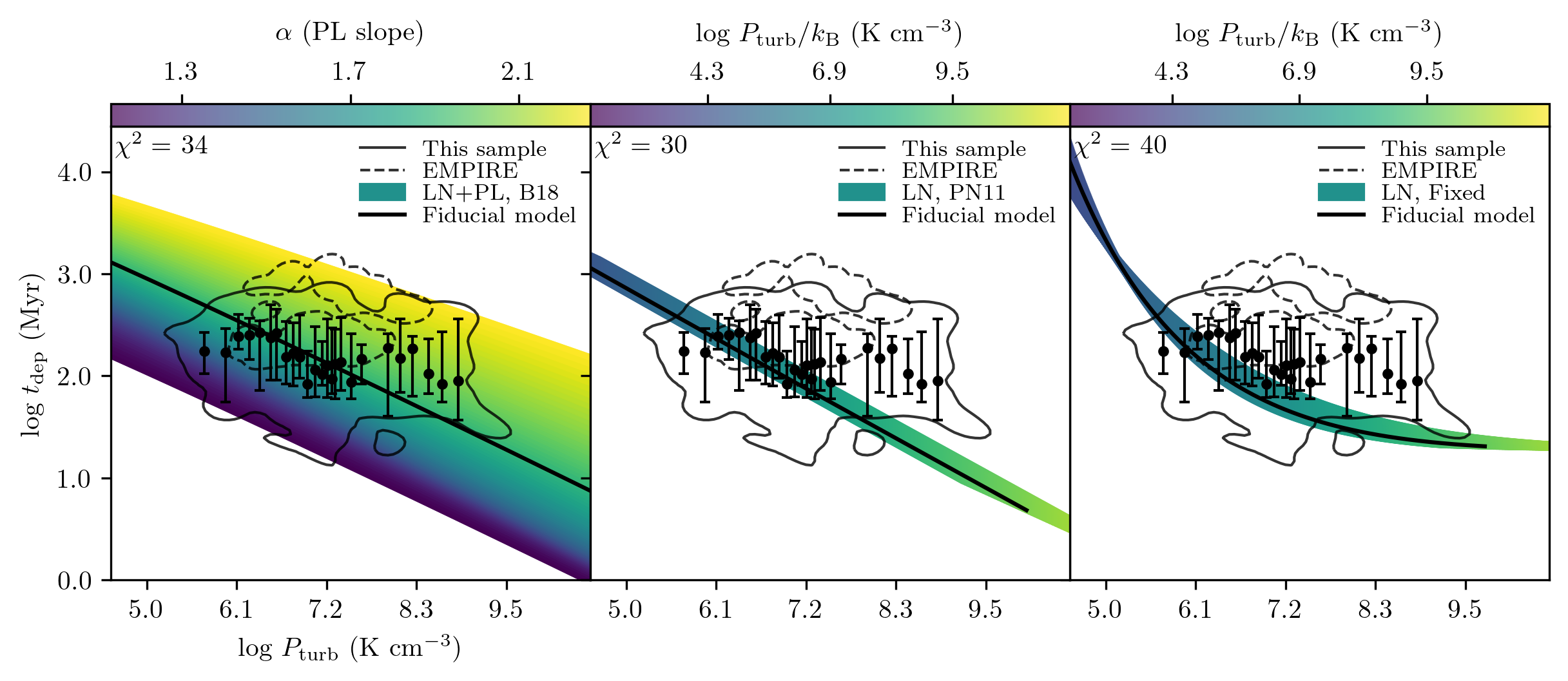}
    \includegraphics[width=0.95\textwidth]{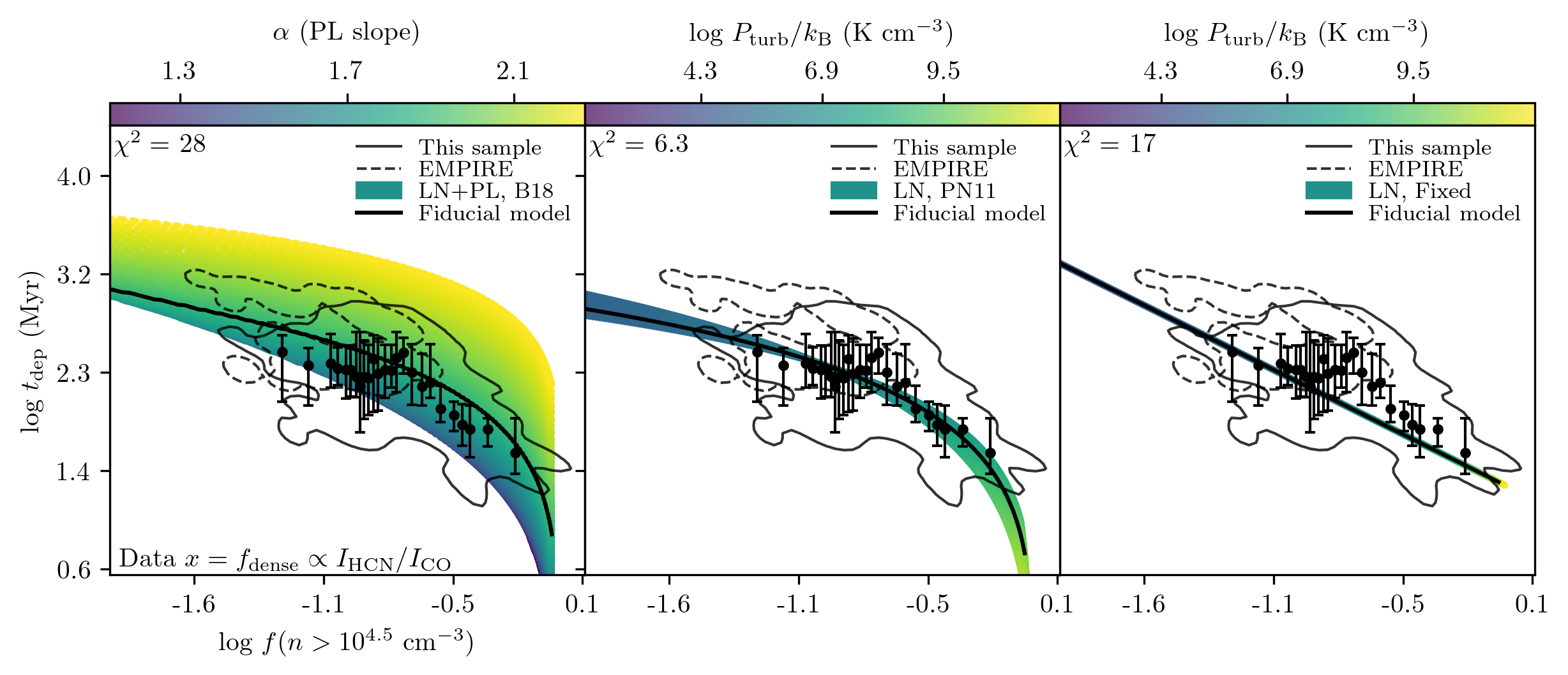}
    \includegraphics[width=0.95\textwidth]{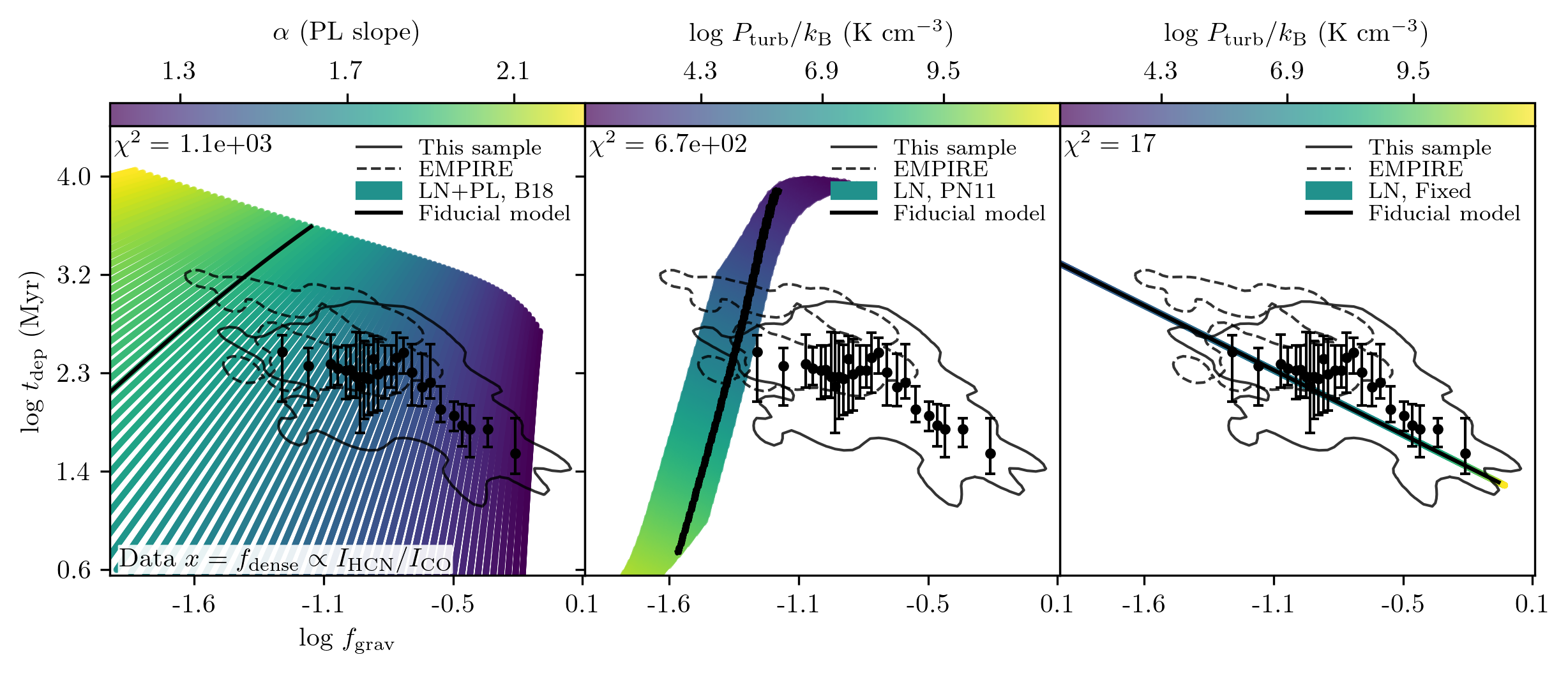}
    \caption{Model $t_\mathrm{dep}$ as a function of $P_\mathrm{turb}$ (top), $f(n>10^{4.5}$ cm$^{-3}$) (middle), and $f_\mathrm{grav}$ (bottom) compared to the data from our sample (solid black contours) and the EMPIRE sample (dashed black contours). See Table \ref{tab:obs_vs_model} for information on how these values are calculated. The LN+PL B18 models are colored by $\alpha_\mathrm{PL}$, and the LN models are colored by $P_\mathrm{turb}$. $\chi^2$ relative to the fiducial model is shown in the upper left corner of each plot. The median values of the data are shown with the black datapoints, and the errorbars correspond to the $1-\sigma$ spread in each bin. Similar to Fig. \ref{fig:eff_vdisp}, the bottom two rows show the importance of distinguishing $f_\mathrm{grav}$ from $f(n>10^{4.5}$ cm$^{-3})$ when using $I_\mathrm{HCN}/I_\mathrm{CO}$ as a tracer of the dense, star forming gas. All models considered agree better with observations when using $f(n>10^{4.5}$ cm$^{-3})$ as a proxy for the dense gas fraction traced by $I_\mathrm{HCN}/I_\mathrm{CO}$. Please see the text for a full analysis of the plots. \label{fig:tdep}}
\end{figure*}

Now we consider if estimates of the total gas depletion times may give insight into the discrepancy between $f_\mathrm{grav}$ and observational estimates of $f_\mathrm{dense}$. We plot $t_\mathrm{dep}$ as a function of $P_\mathrm{turb}$ and $f_\mathrm{dense}$ for our sample of galaxies in Fig. \ref{fig:tdep_data}.  We find moderate negative correlations between $t_\mathrm{dep}$ and the dense gas fraction traced by $I_\mathrm{HCN}/I_\mathrm{CO}$. $t_\mathrm{dep}$ estimated from the data appears relatively constant with $P_\mathrm{turb}$, but the Spearman rank coefficient of the combined data indicates a moderate, negative correlation (cf. Fig. \ref{fig:tdep_data}).

\begin{enumerate}
    \item \textit{LN PN11:} These models are able to reproduce the average trends seen in our data (i.e. negative trend between $t_\mathrm{dep}$ and $P_\mathrm{turb}$ and negative trend between $t_\mathrm{dep}$ and $f(n>10^{4.5}\ \mathrm{cm}^{-3})$). Variations in $P_\mathrm{turb}$ alone do not reproduce the spread seen in our data. $t_\mathrm{dep}$ vs. $f_\mathrm{grav}$ does not track the relationship between $t_\mathrm{dep}$ and $f_\mathrm{dense}$ seen in the data.  
    \item \textit{LN+PL B18:}  We find good agreement between these models and the data when comparing $t_\mathrm{dep}$, $P_\mathrm{turb}$, and $f(n>10^{4.5}\ \mathrm{cm}^{-3})$ (Fig. \ref{fig:tdep}). Although the data overlap with $t_\mathrm{dep}$ vs. $f_\mathrm{grav}$, the models show a much larger spread than that observed in the data.
    \item \textit{LN Fixed:} These models predict that $t_\mathrm{dep}$ decreases until $P_\mathrm{turb}\sim10^7$ K cm$^{-3}$ until it reaches a roughly constant value. This behavior is not seen in our data. $t_\mathrm{dep}$ monotonically decreases with $f(n>10^{4.5}\ \mathrm{cm}^{-3})$ and $f_\mathrm{grav}$ for these models. 
\end{enumerate}

The varying-threshold models best reproduce the observed trend between $t_\mathrm{dep}$ and $P_\mathrm{turb}$. The  varying-threshold models also perform better than the fixed-threshold at reproducing the observed trend between $t_\mathrm{dep}$ and $f_\mathrm{dense}$. Yet again, the trends in the data are best reproduced by the models when considering $f(n>10^{4.5}\ \mathrm{cm}^{-3})$ rather than $f_\mathrm{grav}$.

\subsubsection{Dense Gas Depletion Times}

\begin{figure*}[tb]
    \centering
    \includegraphics[width=0.49\textwidth]{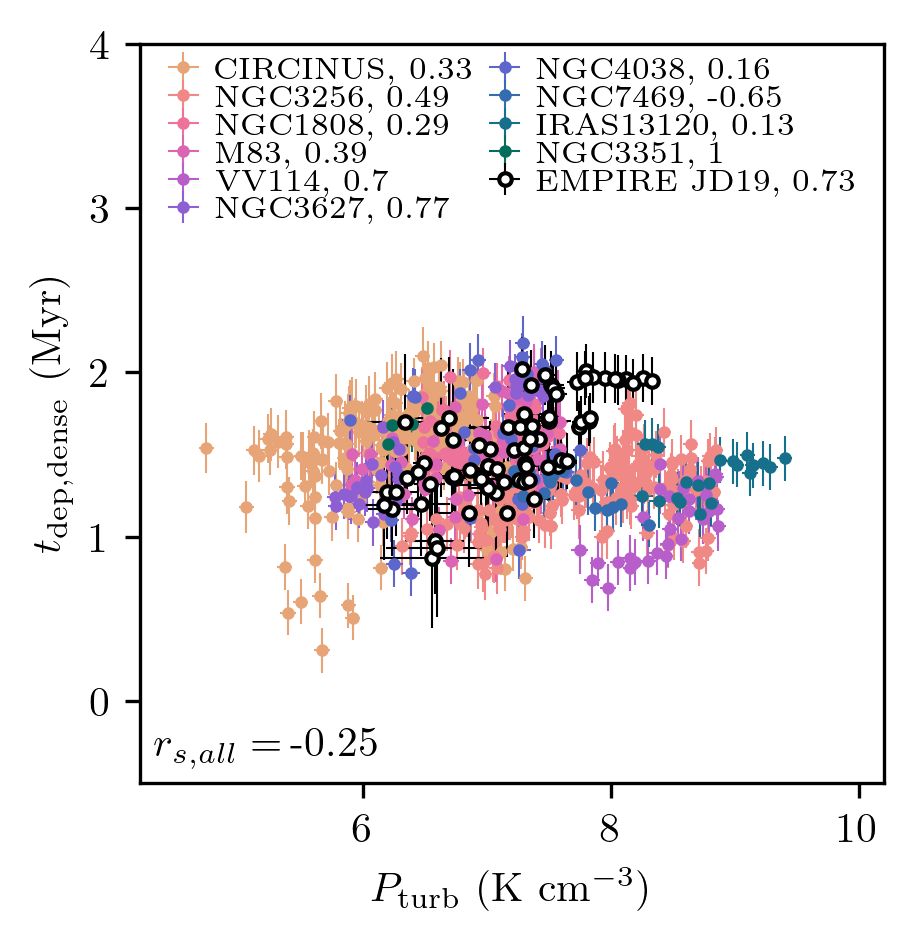}
    \includegraphics[width=0.49\textwidth]{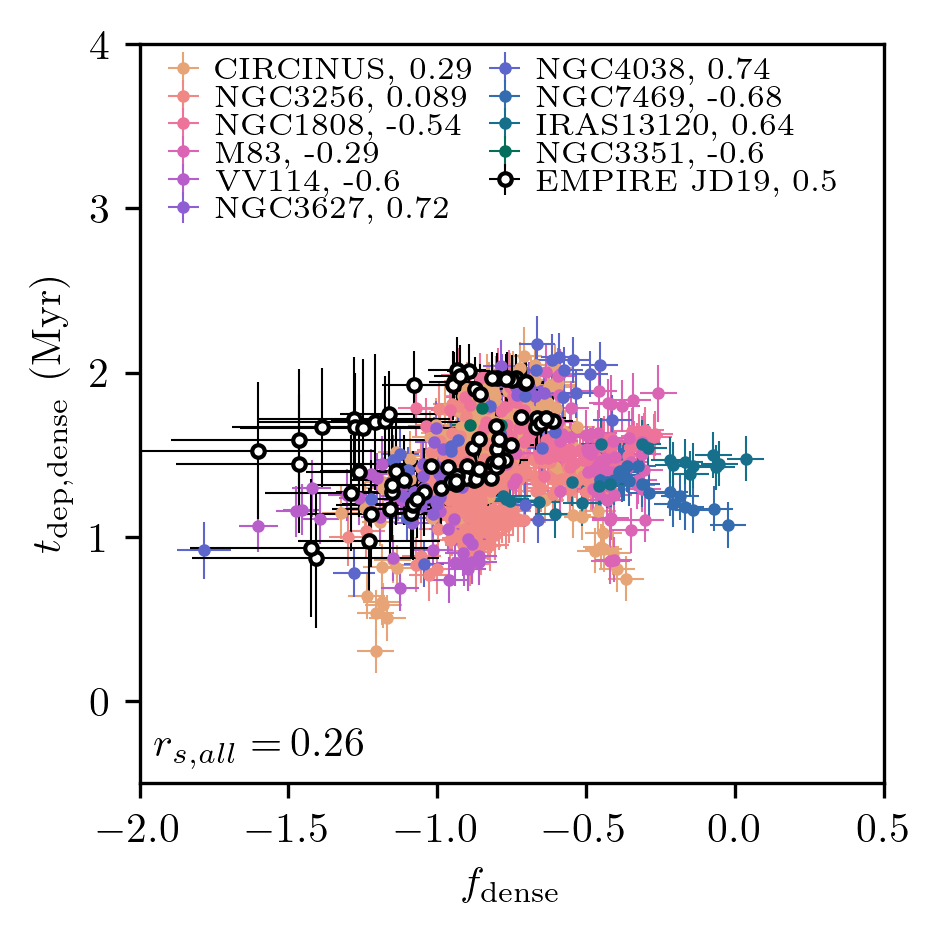}
    \caption{\textit{Left:} Depletion time of the dense molecular gas content as a function of $P_\mathrm{turb}$for our galaxies and the EMPIRE sample. Correlation coefficients are shown and data is colorized by galaxy, similar to the formatting in Fig. \ref{fig:KS_data}. The sample as a whole shows a weak, negative correlation between $t_\mathrm{dep,dense}$ and $P_\mathrm{turb}$. The disk galaxies in the EMPIRE sample show a strong, positive correlation. \textit{Right:}  Depletion time of the dense molecular gas content as a function of $f_\mathrm{dense}$ (right). The sample as a whole shows a weak, positive correlation between $t_\mathrm{dep}$ and $P_\mathrm{turb}$. The disk galaxies in the EMPIRE sample show a moderate, positive correlation. Data is formatted as it is in Fig \ref{fig:KS_data}. \label{fig:tdep_dense_data}}
\end{figure*}

\begin{figure*}[tb]
    \centering
    \includegraphics[width=0.9\textwidth]{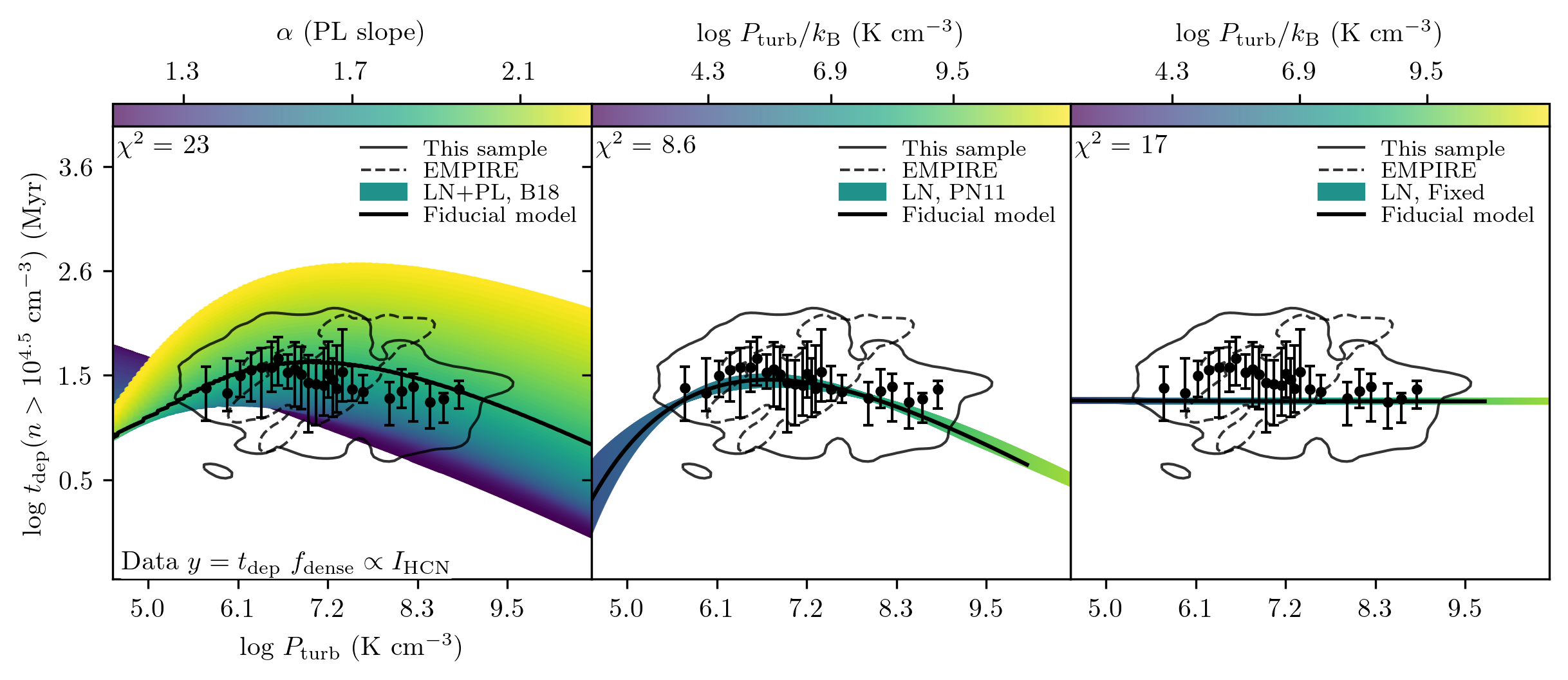}
    \includegraphics[width=0.9\textwidth]{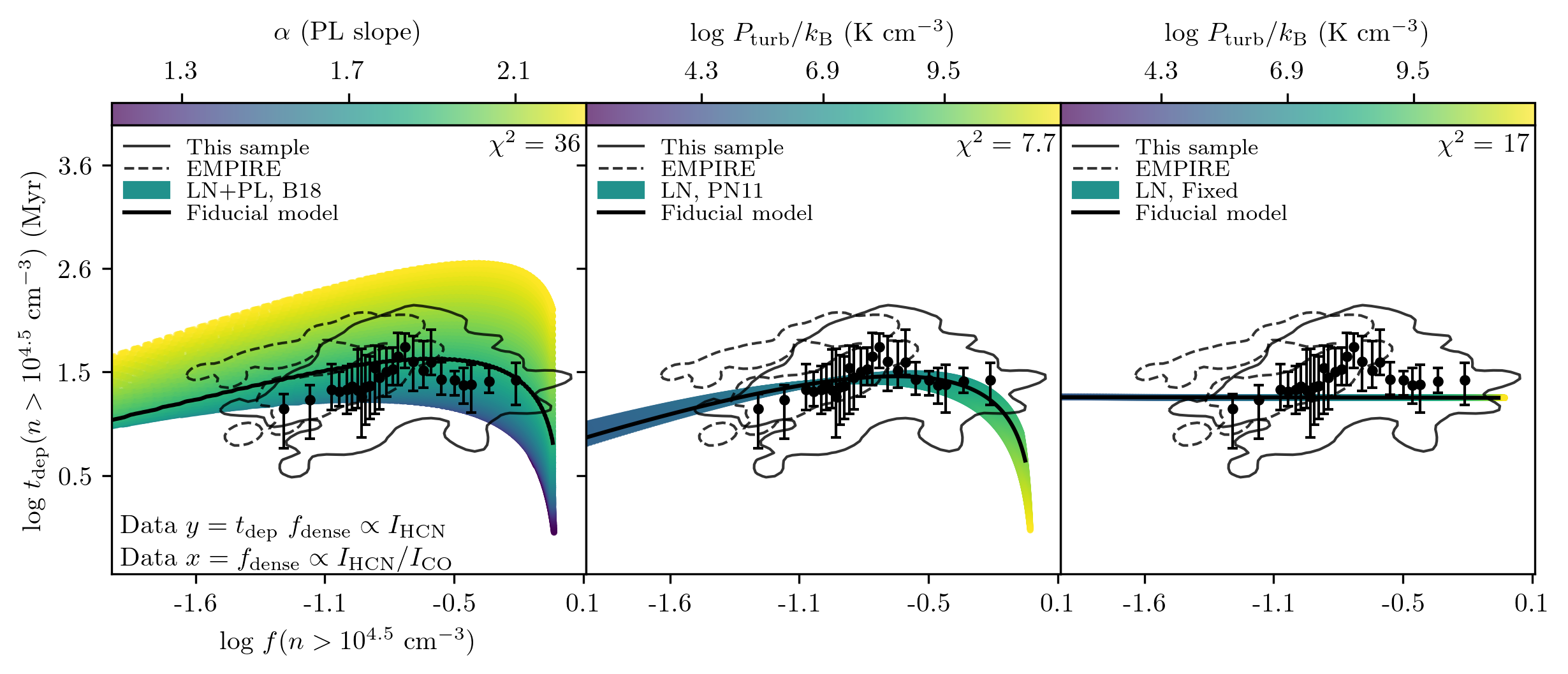}
    \includegraphics[width=0.9\textwidth]{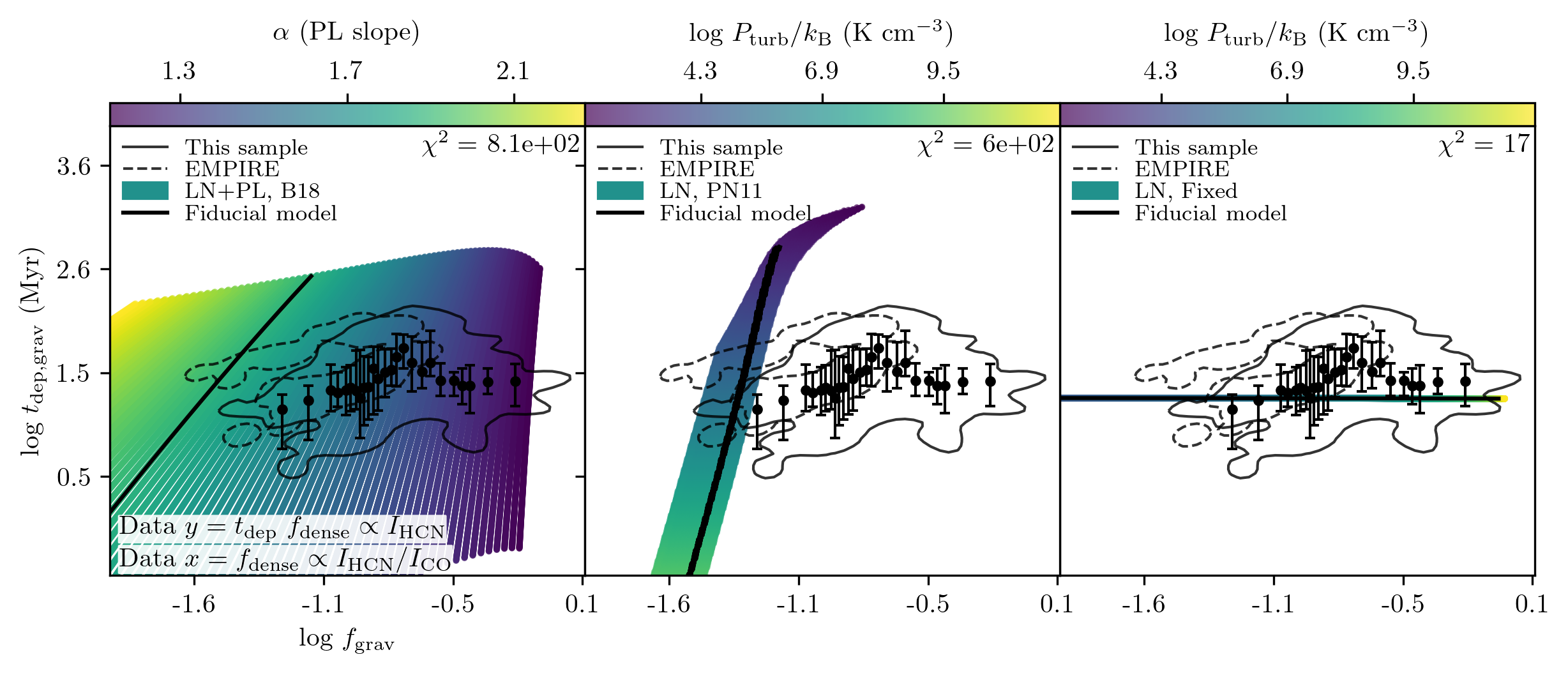}
    \caption{\textit{Top two rows:} Depletion time of the molecular gas above $n = 10^{4.5}$ cm$^{-3}$ as a function of $P_\mathrm{turb}$ (top) and $f(n>10^{4.5}$ cm$^{-3}$) (middle). \textit{Bottom row:} Depletion time of the gravitationally-bound molecular gas as a function of $f_\mathrm{grav}$. These trends are compared against our data (solid black contours) and the EMPIRE sample (dashed black contours). See Table \ref{tab:obs_vs_model} for information on how these values are calculated. The LN+PL B18 models are colored by $\alpha_\mathrm{PL}$, and the LN models are colored by $P_\mathrm{turb}$. $\chi^2$ relative to the fiducial model is shown in the upper left corner of the plots in the top row, and the upper right corner of the plots in the bottom two rows. The median values of the data are shown with the black datapoints, and the errorbars correspond to the $1-\sigma$ spread in each bin. All models considered agree better with observations when using $f(n>10^{4.5}$ cm$^{-3})$ as a proxy for the dense gas fraction traced by $I_\mathrm{HCN}/I_\mathrm{CO}$. Please see the text for a full analysis of the plots. \label{fig:tdep_dense}}
\end{figure*}

We plot the depletion time of the dense gas as traced by HCN, $t_\mathrm{dep,dense}$, as a function of $P_\mathrm{turb}$ (left) and dense gas fraction (right) in Fig. \ref{fig:tdep_dense_data}.  We see a moderate positive correlation between $t_\mathrm{dep,dense}$ and $f_\mathrm{dense}$ in $\sim$ 3 of our sources and the EMPIRE galaxies. The Spearman rank coefficient shows a weak, positive correlation for the combined dataset.  At first this appears counter to the expectation of star formation models, but all models except for those with a fixed threshold are able to roughly reproduce this increase in $t_\mathrm{dep,dense}$ with gas fraction (cf. Fig. \ref{fig:tdep_dense}).

\begin{enumerate}
    \item \textit{LN PN11:} These models predict a turnover in $t_\mathrm{dep}(n>10^{4.5}\ \mathrm{cm}^{-3})$ with $P_\mathrm{turb}$ and in $t_\mathrm{dep}(n>10^{4.5}\ \mathrm{cm}^{-3})$ with $f(n>10^{4.5}\ \mathrm{cm}^{-3})$ (cf. middle column, top and middle rows, respectively Fig. \ref{fig:tdep_dense}). This qualitatively agrees with the average trends seen in our data. We find that $t_\mathrm{dep,grav}$ vs. $f_\mathrm{grav}$ does not track the relationship between $t_\mathrm{dep}$ and $f_\mathrm{dense}$ seen in the data. Variations in $P_\mathrm{turb}$ alone do not reproduce the spread seen in our data.
    \item \textit{LN+PL B18:}  For $\alpha_\mathrm{PL}>1.7$, these models predict a turnover in $t_\mathrm{dep}(n>10^{4.5}\ \mathrm{cm}^{-3})$ around $P_\mathrm{turb}\sim10^7$ K cm$^{-3}$ and for $\mathrm{log}\ f(n>10^{4.5}\ \mathrm{cm}^{-3})\sim-0.5$. For $\alpha_\mathrm{PL}\le 1.7$, $t_\mathrm{dep}(n>10^{4.5}\ \mathrm{cm}^{-3})$ consistently decreases with $P_\mathrm{turb}$ and $f(n>10^{4.5}\ \mathrm{cm}^{-3})$. The observed trends in the data are in qualitative agreement with the trends seen for $\alpha_\mathrm{PL}> 1.7$. We include $t_\mathrm{dep},grav$ vs. $f_\mathrm{grav}$ for completeness, but again find that the models show a much larger spread than that observed in the data.
    \item \textit{LN Fixed:} As predicted in \S \ref{sec:model_predictions}, these models return a fixed $t_\mathrm{dep}(n>10^{4.5}\ \mathrm{cm}^{-3})$ regardless of variations in $P_\mathrm{turb}$ or $f(n>10^{4.5}\ \mathrm{cm}^{-3})$. 
\end{enumerate}

The LN+PL B18 models are able to qualitatively reproduce the trends observed in $t_\mathrm{dep,dense}$, $P_\mathrm{turb}$, and $f_\mathrm{dense}$ for $\alpha_\mathrm{PL}>1.7$. The LN PN11 models are also able to reproduce the average trends, but variations in $P_\mathrm{turb}$ do not reproduce the observed scatter. The above results again provide support against the interpretation of the $I_\mathrm{HCN}/I_\mathrm{CO}$ ratio as a a tracer of $f_\mathrm{grav}$. The observed trends are best reproduced by the varying-threshold models when considering $f(n>10^{4.5})$ cm$^{-3}$, but not with $f_\mathrm{grav}$, and this is why our results provide support against this interpretation.

\section{Discussion}

In this work, we explore how well the HCN-to-CO ratio is tracing the star forming fraction of gas in molecular clouds across the more extreme environments of U/LIRGs, mergers, and galaxy centers using gravoturbulent models of star formation \citep{Burkhart:2018,Burkhart:2019,Padoan:2011,Krumholz:2005,Federrath:2012,Hennebelle:2011}. Previous studies find that HCN may be tracing gas above a fairly constant density and that the mass traced by HCN appears to scale linearly with the SFR, indicating a constant, average depletion time of the dense gas that spans clouds \citep[cf.][]{Wu:2005} to entire galaxies including disk galaxies and U/LIRGs \citep[cf.][]{Gao:2004a,Gao:2004b}. This result is consistent with the predictions of the LN-only models with a fixed threshold that we consider in our analysis (see the right column of Figs. \ref{fig:KS_plots}, \ref{fig:gas_fractions}, \ref{fig:eff_vdisp}, \ref{fig:tdep}, and \ref{fig:tdep_dense}). These models are able to qualitatively reproduce some of the average trends involving the dense, molecular gas traced by HCN such as the decrease in $t_\mathrm{dep}$ with increasing $f_\mathrm{dense}\propto{I_\mathrm{HCN}/I_\mathrm{CO}}$ (Fig. \ref{fig:tdep}). However, these models fail to capture the spread observed in many of the trends of our data and as well as the observed trends with $P_\mathrm{turb}$.
\par
The behavior of star formation laws observed in disk galaxies may give some insight into the discrepancy between the fixed-threshold models and observations. There is evidence of non-linear scalings between dense gas mass and the SFR within disk galaxies that appears to be correlated with galactic radius in disk galaxies \citep[e.g.][]{Gallagher:2018,Usero:2015,Chen:2015}. Relative to their disks, the centers of late-type galaxies show longer dense gas depletion times despite larger dense gas fractions traced by $I_\mathrm{HCN}/I_\mathrm{CO}$ \citep[e.g.][]{Gallagher:2018,Usero:2015,Chen:2015}. Similarly, the nuclei of the merging Antennae galaxies also display longer dense gas depletion times despite higher dense gas fractions compared to the Overlap region of this merging system \citep{Bemis:2019,Bigiel:2015}. Ambient pressure of the ISM is higher in the centers of disk galaxies compared to larger radii, which is apparent from observational estimates of $P_\mathrm{turb}$ from CO across the disks of nearby disk galaxies \citep[cf.][]{Gallagher:2018}. If higher $P_\mathrm{turb}$ is indicative of higher turbulent support, then gravoturbulent models of star formation with varying star formation thresholds may offer an explanation for these observed trends \citep[i.e.][]{Burkhart:2018,Padoan:2011,Krumholz:2005}.
\par
The varying-threshold models considered in this paper \citep[i.e.][]{Burkhart:2018,Padoan:2011} predict that the gas density required for the onset of collapse (and therefore star formation) increases in the presence of supportive processes such as solenoidal turbulence. This is apparent in Fig. \ref{fig:gas_fractions} (left two columns, top row) where the LN+PL B18 and LN P11 models show a decrease in $f_\mathrm{grav}$ with increasing $P_\mathrm{turb}$. Turbulence can also increase the mean gas density of the molecular cloud as a whole by widening the $n-$PDF. This is also apparent in Fig. \ref{fig:gas_fractions} (left two columns, bottom row) where the LN+PL B18 and LN P11 models show an \textit{increase} in $f(n>10^{4.5}$ cm$^{-3})$ with increasing $P_\mathrm{turb}$. The difference in behavior between $f_\mathrm{grav}$ and $f(n>10^{4.5}$ cm$^{-3})$ with $P_\mathrm{turb}$ reflects an important prediction of these varying-threshold models: the supportive effect of turbulence can have a more significant impact on star formation than the increase in dense gas mass due to the widening of the $n-$PDF.
\par
A wider $n-$PDF would naturally result in enhanced $I_\mathrm{HCN}/I_\mathrm{CO}$ if this ratio is tracing gas above a roughly constant density \citep[cf.][]{Leroy:2017a,Bemis:2020}. At the same time, the star formation rate itself can appear suppressed relative to the total dense gas mass because the onset of star formation occurs at a higher density than the mean density of the gas traced by $I_\mathrm{HCN}/I_\mathrm{CO}$. The results of our work support this picture. In particular, the LN+PL B18 and LN P11 models are able to qualitatively reproduce the average trends observed in our data when considering $f(n>10^{4.5}$ cm$^{-3}$) as a proxy for the dense gas fraction traced by $I_\mathrm{HCN}/I_\mathrm{CO}$ (cf. Figs. \ref{fig:gas_fractions}, \ref{fig:eff_vdisp}, \ref{fig:tdep}, and \ref{fig:tdep_dense}). 
\par
One apparent discrepancy between the data and the varying-threshold models is the behavior of $\epsilon_\mathrm{ff}$ with $P_\mathrm{turb}$ (cf. top row, Fig. \ref{fig:eff_vdisp}). Our sample shows a weak, negative trend between $\epsilon_\mathrm{ff}$ and $P_\mathrm{turb}$ and the EMPIRE sample has a moderate, negative trend. Opposite to this, the varying-threshold models predict an increase in $\epsilon_\mathrm{ff}$ with $P_\mathrm{turb}$ (cf. Fig. \ref{fig:eff_vdisp}). The fixed-threshold LN model predicts that $\epsilon_\mathrm{ff}$ initially increases with $P_\mathrm{turb}$ and then decreases towards higher pressures. The discrepancy between the models and data is likely in part due to inaccurate estimates of $t_\mathrm{ff}$ from our data. When estimating the mean density of the molecular gas traced by CO, we assume a fixed line-of-sight depth of 100 pc for all measurements, including those from the galaxies in the EMPIRE sample. We can also see from Fig. \ref{fig:tdep} that the models predict a steeper decline in $t_\mathrm{dep}$ with $P_\mathrm{turb}$ than what is observed in the data. These two things combined likely contribute to the discrepancy between the data and the models when considering $\epsilon_\mathrm{ff}$ as a function of $P_\mathrm{turb}$. Due to the weakness of the observed trend in our data and the uncertainty in $t_\mathrm{ff}$, it is difficult to determine if $\epsilon_\mathrm{ff}$ truly increases or decreases with $P_\mathrm{turb}$. The apparent agreement between $\epsilon_\mathrm{ff}$ and $f(n>10^{4.5}\  \mathrm{cm}^{-3})$ predicted by the varying threshold models and our data may then be a reflection of the stronger connection between star formation and denser gas that is better traced by HCN relative to the bulk molecular gas traced by CO \citep[cf.][]{Gao:2004a,Gao:2004b}.
\par
We also find that the scatter of the data is not well-reproduced by variations in $P_\mathrm{turb}$ when considering the LN-only models (e.g. Figs. \ref{fig:gas_fractions}, \ref{fig:eff_vdisp}, \ref{fig:tdep}, and \ref{fig:tdep_dense}). One parameter that we did not vary in this analysis is $\alpha_\mathrm{vir}$. We consider here how this would impact our results. For the \citet{Padoan:2011} models, higher $\alpha_\mathrm{vir}$ increases $n_\mathrm{SF}$ such that more unbound clouds will have lower $\epsilon_\mathrm{ff}$. Variations in $\alpha_\mathrm{vir}$ would introduce some scatter into the results, primarily in the relationships that depend on $\epsilon_\mathrm{ff}$ and $t_\mathrm{dep}$. However, variations in $\mathcal{M}$ have an impact on both the width of the $n-$PDF (Eq. \ref{eq:sigma_s_chap3}) and the shift in $n_\mathrm{SF}$ (Eq. \ref{eq:ncrit_km05}) for the LN PN11 models, while changes in $\alpha_\mathrm{vir}$ primarily have an impact on $n_\mathrm{SF}$. Changes in the dense gas fraction indicated by $f(n>10^{4.5}$ cm$^{-3}$) are not analytically dependent on $\alpha_\mathrm{vir}$, and therefore would need an additional explanation. 
\par
The LN+PL B18 models offer an alternative explanation for the scatter of the data: variations in PL slope ($\alpha_\mathrm{PL}$). Simulations show that a PL tail develops as a consequence of cloud evolution as gas becomes self-gravitating, and the slope of the PL tail becomes more shallow as the fraction of bound gas (i.e. $f_\mathrm{grav}$) increases \citep[cf. ][]{Ballesteros-Paredes:2011a,Collins:2012,Schneider:2015a,Burkhart:2017,Padoan:2017}. This is supported by observations of some Milky Way clouds which show a PL tail at high densities \citep[e.g.][]{Kainulainen:2009,Schneider:2013,Lombardi:2015,Schneider:2015a,Schneider:2016,Alves:2017}. Additionally, there is observational evidence for a connection between PL slope and $\epsilon_\mathrm{ff}$ which matches the predictions of the LN+PL B18 models \citep{Burkhart:2019,Federrath:2013}.
\par
The scatter in our data is well-reproduced for $\alpha_\mathrm{PL}=1-2.3$ (see the left column of Figs. \ref{fig:KS_plots}, \ref{fig:gas_fractions}, \ref{fig:eff_vdisp}, \ref{fig:tdep}, and \ref{fig:tdep_dense}). These models are also able to reproduce the average trends of our data. Out of the three models we consider, the LN+PL B18 models perform best at reproducing the observational trends and scatter of our data and the EMPIRE sample. In the context of the LN+PL B18 models, the scatter in our data is therefore driven in part due to cloud evolution \citep{Ballesteros-Paredes:2011a}, such that the $n-$PDF of clouds evolves from a lognormal shape to a composite lognormal and powerlaw over time as more gas becomes gravitationally-bound. The underlying trends are still driven largely by variations in turbulence, which we impose in our model parameter space by using the observational relationship between gas surface density and velocity dispersion found by \citet{Sun:2018} (cf. Eq. \ref{eq:sun18}). 

\section{Conclusions}

We find the following conclusions from our comparison between observation and the predictions of analytical models of star formation:
\begin{enumerate}
    \item  KS relationship: For the model parameter space considered, none of the models reproduce the multi-slope KS relationship as found in \citet[][]{Wilson:2019}. Variations in the slope of the  $\Sigma_\mathrm{mol}-\sigma_v$ relationship can produce variations in the KS slope, such that a shallower (steeper) $\Sigma_\mathrm{mol}-\sigma_v$ relationship would produce a shallower (steeper) KS relationship. Furthermore, systematic changes in $\epsilon_0$ or PL slope with $\Sigma_\mathrm{mol}$ could also affect the KS slope.  Increasing (decreasing) PL slope with $\Sigma_\mathrm{mol}$ would produce a shallower (steeper) KS relationship. Similarly, decreasing (increasing) $\epsilon_0$ with $\Sigma_\mathrm{mol}$ produces a shallower (steeper) KS slope.
    \item $f_\mathrm{dense}$ and $f_\mathrm{grav}$: For the PN11 and B18 models, $f_\mathrm{grav}$ decreases with $P_\mathrm{turb}$, and increases for the fixed-threshold models. $f(n>10^{4.5}\ \mathrm{cm}^{-3})$ increases with $P_\mathrm{turb}$ for all models. The behavior of $f(n>10^{4.5}\ \mathrm{cm}^{-3})$ with $P_\mathrm{turb}$ is in qualitative agreement with our data, which shows an increase in $f_\mathrm{dense}=(\alpha_\mathrm{HCN}/\alpha_\mathrm{CO}) I_\mathrm{HCN}/I_\mathrm{CO}$ with observational estimates of $P_\mathrm{turb}$. We therefore conclude that the $I_\mathrm{HCN}/I_\mathrm{CO}$ ratio likely does not track $f_\mathrm{grav}$ as predicted by the varying-threshold models, but rather the fraction of gas above some fixed density, such as $f(n>10^{4.5}\ \mathrm{cm}^{-3})$.
    \item $\epsilon_\mathrm{ff}$: For the PN11 and B18 models, $\epsilon_\mathrm{ff}$ increases with $P_\mathrm{turb}$ on average. For the fixed-threshold models, $\epsilon_\mathrm{ff}$ increases with $P_\mathrm{turb}$ until $P_\mathrm{turb}/k_\mathrm{B}\approx10^{7.5}\ \mathrm{K\ cm}^{-3}$ where it turns over. All models predict an increase in $\epsilon_\mathrm{ff}$ with $f(n>10^{4.5}\ \mathrm{cm}^{-3})$, on average, with only the fixed-threshold models showing a turnover in this relationshop at high fractions. $\epsilon_\mathrm{ff}$ decreases with $f_\mathrm{grav}$ for the the PN11 models and the B18 models for fixed $\alpha_\mathrm{PL}$. For the fixed-threshold models, $\epsilon_\mathrm{ff}$ increases with $f_\mathrm{grav}$ until it turns over at high fractions. Under the assumption of a fixed l.o.s. depth, our data show a weak decrease in $\epsilon_\mathrm{ff}$ with $P_\mathrm{turb}$ and increase with $f_\mathrm{dense}$. This is in qualitative agreement with the PN11 and B18 models, assuming $I_\mathrm{HCN}/I_\mathrm{CO}$ is tracking $f(n>10^{4.5}\ \mathrm{cm}^{-3})$, but not $f_\mathrm{grav}$.
    \item Total gas depletion time: All models predict a decrease of $t_\mathrm{dep}$ with both $P_\mathrm{turb}$ and $f(n>10^{4.5}\ \mathrm{cm}^{-3})$. The fixed-threshold models predict a flattening of $t_\mathrm{dep}$ with higher $P_\mathrm{turb}$ but a constant decrease with $f(n>10^{4.5}\ \mathrm{cm}^{-3})$. Contrary to this, the PN11 models and the B18 models show a constant decrease in $t_\mathrm{dep}$ with $P_\mathrm{turb}$, but a steepening decline in $t_\mathrm{dep}$ with higher $f(n>10^{4.5}\ \mathrm{cm}^{-3})$. $t_\mathrm{dep}$ increases with  $f_\mathrm{grav}$ for the PN11 models and the B18 models for fixed $\alpha_\mathrm{PL}$. The trends in data are in better qualitative agreement with the PN11 and B18 models, again assuming $I_\mathrm{HCN}/I_\mathrm{CO}$ is tracking $f(n>10^{4.5}\ \mathrm{cm}^{-3})$, but not $f_\mathrm{grav}$.
    \item Dense gas depletion time: The PN11  and B18 models (for $\alpha_\mathrm{PL}>1.7$) predict a turnover in the depletion time of dense gas, $t_\mathrm{dep}(n>10^{4.5}\ \mathrm{cm}^{-3})$, with $P_\mathrm{turb}$ and $f(n>10^{4.5}\ \mathrm{cm}^{-3})$, while the fixed threshold models predict a constant dense gas depletion time \citep{Gao:2004a,Gao:2004b}. We note that $\alpha_\mathrm{PL}<1.7$ produce a nearly constant decrease in $t_\mathrm{dep}(n>10^{4.5}\ \mathrm{cm}^{-3})$ with $P_\mathrm{turb}$ and $f(n>10^{4.5}\ \mathrm{cm}^{-3})$. $t_\mathrm{dep,grav}$ increases with  $f_\mathrm{grav}$ for the PN11 models and the B18 models for fixed $\alpha_\mathrm{PL}$. The trends in data are in better qualitative agreement with the PN11 and B18 models, again assuming $I_\mathrm{HCN}/I_\mathrm{CO}$ is tracking $f(n>10^{4.5}\ \mathrm{cm}^{-3})$, but not $f_\mathrm{grav}$.
    \item Variations in PL slope ($\alpha_\mathrm{PL}$) (lognormal+powerlaw B18 models) are able to explain the scatter in the data. The scatter in our data is not well-reproduced by variations in $P_\mathrm{turb}$ (or $\mathcal{M}$) alone.
    \item The varying-threshold models (lognormal+powerlaw B18 and lognormal-only PN11) are able to better reproduce the average trends of our data and the EMPIRE sample compared to the fixed-threshold lognormal-only models.
    \item The average trends in the data with $P_\mathrm{turb}$ and $f_\mathrm{dense}$ are likely set by variations in turbulence determined by local environment within each galaxy.
\end{enumerate}

\noindent In summary, the HCN/CO ratio is likely a reliable tracer of gas above a constant density, such as $n_\mathrm{SF}\approx10^{4.5}$ cm$^{-3}$, but not necessarily $f_\mathrm{grav}$. If varying thresholds for star forming gas  exist in nature, then the HCN/CO ratio is a poor tracer of $f_\mathrm{grav}$. 

\section{Acknowledgements}

This paper makes use of the following ALMA data: ADS/JAO.ALMA\#2011.0.00467.S, \linebreak
ADS/JAO.ALMA\#2011.0.00525.S, \linebreak
ADS/JAO.ALMA\#2011.0.00772.S, \linebreak
ADS/JAO.ALMA\#2012.1.00165.S, \linebreak
ADS/JAO.ALMA\#2012.1.00185.S, \linebreak
ADS/JAO.ALMA\#2012.1.01004.S, \linebreak
ADS/JAO.ALMA\#2013.1.00218.S, \linebreak
ADS/JAO.ALMA\#2013.1.00247.S, \linebreak
ADS/JAO.ALMA\#2013.1.00634.S, \linebreak
ADS/JAO.ALMA\#2013.1.00885.S, \linebreak
ADS/JAO.ALMA\#2013.1.00911.S, \linebreak
ADS/JAO.ALMA\#2013.1.01057.S, \linebreak
ADS/JAO.ALMA\#2015.1.00993.S, \linebreak
ADS/JAO.ALMA\#2015.1.01177.S, \linebreak
ADS/JAO.ALMA\#2015.1.01286.S, \linebreak
ADS/JAO.ALMA\#2015.1.01538.S. ALMA is a partnership of ESO (representing its member states), NSF (USA) and NINS (Japan), together with NRC (Canada), MOST and ASIAA (Taiwan), and KASI (Republic of Korea), in cooperation with the Republic of Chile. The Joint ALMA Observatory is operated by ESO, AUI/NRAO and NAOJ. 

\software{This work made use of the following software packages:  \textsc{astropy} \citep{astropy:2013,astropy:2018,astropy:2022}, \textsc{casa} \citep{McMullin:2007}, \textsc{cmasher} \citep{Vandervelden:2020}, \textsc{matplotlib} \citep{Hunter:2007}, \textsc{numpy} \citep{Harris:2020}, \textsc{scipy} \citep{Virtanen:2020}, \textsc{spectral-cube} (\url{https://spectral-cube.readthedocs.io/en/latest/}).}

\clearpage
\appendix

\section{Appendix: Moment Maps and Uncertainties} \label{ap:uncertainties}

We use the following expressions to calculate moments and their corresponding uncertainties.
\begin{align}
    {\textsc{Moment 0}:\qquad}
        I &=  \Sigma_i T_i \Delta v \\[0.5em]
        \delta I &= \delta \bar{T}\Delta v \sqrt{N_\mathrm{chan}} \label{eq:unc_mom0} \\[0.5em]
    {\textsc{Moment 1}:\qquad}
        \bar{v} &= \frac{\Sigma_i T_i v_i}{\Sigma_i T_i}\label{eq:mean_vel}  \\[0.5em] 
        \delta{\bar{v}} &= \frac{N_\mathrm{chan}\Delta v}{2\sqrt{3}} \frac{\delta I}{I}\label{eq:unc_mom1}  \\[0.5em]
    {\textsc{Moment 2}:\qquad}
        \sigma_v &= \sqrt{\frac{\Sigma_i T_i (v_i - \bar{v})^2}{\Sigma_i T_i}}\label{eq:vel_disp} \\[0.5em]
        \delta \sigma_v &= \frac{\delta_I}{I}\frac{(N_\mathrm{chan}\Delta v)^2}{8\sqrt{5}}\frac{1}{\sigma_v}\label{eq:unc_mom2}
\end{align}
\noindent A full derivation of the uncertainties on Eqs. \ref{eq:unc_mom1} and \ref{eq:unc_mom2} is given in \citet{Wilson:2022}.

\bibliography{refs}{}
\bibliographystyle{aasjournal}

\end{document}